\documentclass[11pt]{article}
\usepackage{tikz} 
\usetikzlibrary{positioning}
\usepackage{geometry}
\usepackage{booktabs}
\usepackage{lineno}
\usepackage{multirow}
\usepackage{enumitem}
\usepackage{amsmath}
\usepackage{float}
\usepackage{tabularx}
\usepackage{graphicx} 
\usepackage{amssymb,amscd,amsmath,amsthm,color}
\usepackage[T1]{fontenc} 
\usepackage{subcaption}
\usepackage{enumitem}
\usepackage{mathtools}
\usepackage{tcolorbox}
\usepackage{verbatim}
\usepackage{mathrsfs}
\usepackage{physics}
\usepackage{mathtools}
\usepackage{slashed}
\usepackage{longtable}
\usepackage{jheppub}
\newcommand{\ret}{{\rm ret}}

\newcommand{\chihat}{\hat{\chi}}
\newcommand{\dt}{\partial_t}

\def\be{\begin{equation}}
\def\ee{\end{equation}}
\def\bea{\begin{eqnarray}}
\def\eea{\end{eqnarray}}

\newcommand{\p}{\partial}

\newcommand{\Scri}{\mathcal{I}^{+}}
\newcommand{\nhat}{\widehat{n}}

\newcommand{\STF}{\mathrm{STF}}

\numberwithin{equation}{section}

\newcommand{\Qrho}{Q^{(\rho)}}
\newcommand{\Qp}{Q^{(p)}}
\newcommand{\Qrp}{Q^{(\rho+p)}}

\makeatletter
\g@addto@macro\bfseries{\boldmath}\g@addto@macro\bfseries{\boldmath}
\makeatother
\title{Octupolar Gravitational Radiation in de Sitter Spacetime}
\author{Harsh$^a$,
Omkar Shetye$^{b}$,  Amitabh Virmani$^a$}
\emailAdd{harsh22@cmi.ac.in, omkar.sanjay.shetye@uj.edu.pl, avirmani@cmi.ac.in}
\affiliation[a]{Chennai Mathematical Institute, 
H1, SIPCOT IT Park, Siruseri, Kelambakkam 603103 India}
\affiliation[b]{ Instytut Fizyki Teoretycznej,
  Uniwersytet Jagiello\'nski,
  ul. \L{}ojasiewicza 11,
  30-348 Krak\'ow,
  Poland
}

\abstract{We derive the gravitational radiation generated by a localised matter source in de Sitter spacetime at octupolar order in the multipolar expansion, working in generalised harmonic gauge in the future Poincar\'e patch. The scalar, vector, and tensor perturbation equations are solved explicitly, taking into account the cosmological tail. This constitutes the first extension of the de Sitter multipolar expansion beyond quadrupolar order. A key technical ingredient is a universal multipole moment differentiation identity, which unifies the relations between multipole moments into a single recursive formula and enables a controlled extension to higher multipoles. Using the symmetric trace-free tensor formalism of Blanchet, Damour, and Iyer, we further develop a systematic translation of the radiative solution into a spherical harmonic basis, and apply it to both the quadrupolar and octupolar metric perturbations. This representation renders the angular structure of the radiation manifest, simplifies the verification of the linearised field equations, and facilitates direct comparison with Bondi-gauge and other perturbative treatments of de Sitter spacetime.}

\makeatletter
\gdef\@fpheader{}
\makeatother
\begin{document}
\allowdisplaybreaks

\maketitle

\section{Introduction}
The accurate modelling of gravitational radiation on cosmological scales is a
problem of growing interest. Third-generation detectors such as the Einstein
Telescope and Cosmic Explorer will observe binary merger events at redshifts of
order ten and beyond, where the background curvature of the universe is no
longer strictly negligible. Since, within $\Lambda$CDM, our universe asymptotes
at late times to de~Sitter spacetime, de~Sitter space is the natural arena in
which to extend the multipolar post-Newtonian and post-Minkowskian framework
developed for asymptotically flat spacetimes. Although the corrections to the emitted power induced by a positive
cosmological constant are suppressed by the minuscule ratio of the dynamical
time scale of the source to the Hubble time~\cite{Ashtekar:2015lxa}, the
primary interest of the problem is conceptual: a gravitational radiation
theory that cannot accommodate $\Lambda > 0$, however small, is incomplete in
a way that deserves to be understood.

The non-triviality of this extension was made sharp by Ashtekar, Bonga, and
Kesavan (ABK)~\cite{Ashtekar:2015lxa}: even the simplest step of generalising
Einstein's 1918 quadrupole formula to include a positive cosmological constant
$\Lambda$ faces several conceptual obstacles. The standard $1/r$ expansions
break down, the Weyl tensor component $\Psi_4^0$ that is routinely used in
numerical simulations acquires an ambiguity due to the spacelike nature of
conformal infinity, and the $H \to 0$ limit of physical quantities can be
discontinuous. A systematic treatment requires a fresh approach to the
linearised field equations. ABK decoupled these equations in a generalised
harmonic gauge~\cite{Ashtekar:2015lxa, deVega:1998ia} and derived a quadrupolar formula for
gravitational radiation in de~Sitter spacetime.

A particularly comprehensive treatment of the quadrupolar radiation in de~Sitter  was given by
Comp\`ere, Hoque, and Kutluk (CHK)~\cite{Compere:2023ktn}, which we take as our
starting point. CHK derived the metric perturbation at quadrupolar order in both
generalised harmonic gauge and Bondi gauge, including both parities, the even
(mass quadrupole) and odd (current quadrupole) sectors. In parallel, Bonga,
Bunster, and P\'erez (BBP)~\cite{Bonga:2023eml} studied gravitational radiation
with $\Lambda > 0$ from a Bondi--Sachs perspective. These papers introduced
boundary conditions adapted to the de~Sitter setting and derived the associated
energy and angular momentum fluxes. In earlier work, two of the present
authors~\cite{Harsh:2024kcl} introduced de~Sitter Teukolsky waves as a natural
basis for linearised quadrupolar gravitational waves in de~Sitter spacetime and
established their relation to the CHK and BBP solutions. More recently, Poole,
Skenderis, and Taylor~\cite{Poole:2025cmv} have presented a comprehensive
covariant phase space treatment of gravitational charges and radiation in
asymptotically locally de~Sitter spacetimes, which also serves as
an excellent guide to the current status of the subject. By now, the quadrupolar problem in
de~Sitter spacetime is well understood, and the stage is set for a
systematic extension to higher multipole orders.

The present paper advances this programme in two principal directions. First,
we extend the multipolar expansion to octupolar order, deriving the complete
metric perturbation in generalised harmonic gauge at this order. A key
technical ingredient that makes the extension tractable is a universal
multipole moment differentiation identity, established inductively in
Appendix~\ref{proof_identity}. This identity unifies into a single recursive
formula the many relations between multipole moments that appear separately in
CHK, reveals their common algebraic origin, and enables a controlled extension
to arbitrary multipole order. The results obtained here provide evidence that
a systematic multipolar hierarchy exists beyond quadrupolar order and may
ultimately support a genuine post-de~Sitter radiation formalism.

Second, we develop a systematic translation of the metric perturbation from the
symmetric trace-free (STF) tensor language, in which the solution to the field
equations naturally emerges, into a spherical harmonic basis. The need for such
a translation is intrinsic to the de~Sitter setting. Unlike Minkowski
spacetime, de~Sitter admits no single coordinate patch in which time-translation
and spatial-translation symmetries are simultaneously manifest: the former is
adapted to the static patch, the latter to the Poincar\'e patch. Results
derived in one description must therefore be systematically recast to interface
with analyses carried out in another, and the spherical harmonic decomposition
provides the natural bridge. The translation is built on the STF formalism of
Blanchet, Damour, and Iyer~\cite{Blanchet:1985sp, Damour:1990gj} and is
developed in detail in Appendices~\ref{app:B}--\ref{app:D}. We apply this
framework to rewrite both the quadrupolar and octupolar metric perturbations
explicitly in terms of scalar, vector, and tensor spherical harmonics. This
decomposition makes the angular structure of the radiation manifest and places
the results in a form suited for direct comparison with the broader literature
on gravitational perturbations in de~Sitter spacetime, including the work of
BBP and the de~Sitter Teukolsky waves mentioned above. It also renders the
verification of the linearised Einstein equations comparatively
straightforward.

To place the present work within a broader context, it is useful to briefly review recent developments concerning gravitational radiation and conserved charges in spacetimes with a positive cosmological constant. Unlike the asymptotically flat or anti-de Sitter cases, conformal infinity in de~Sitter spacetime is spacelike, there is no globally timelike Killing vector, and the standard machinery associated with null or spatial infinity does not carry over directly. These features complicate the formulation of gravitational radiation, the definition of appropriate boundary conditions, and the construction of physically meaningful charges and fluxes. Considerable progress has nevertheless been made in recent years. Important questions include the formulation of a suitable no-incoming-radiation condition and the characterisation of energy carried by gravitational waves in the presence of a positive cosmological constant \cite{Ashtekar:2015lxa, Date:2015kma, Date:2016uzr, Hoque:2018byx,  Ashtekar:2019khv,  Dobkowski-Rylko:2022dva, Compere:2023ktn, Dobkowski-Rylko:2024jmh, Compere:2024ekl}. Closely related investigations have analysed the configurations, energy, and symplectic structure of linearised gravitational fields on de~Sitter backgrounds \cite{Bishop:2015kay, Ashtekar:2015lla, Hoque:2018byx, Chrusciel:2020rlz, Kolanowski:2020wfg, Chrusciel:2021ttc, Chrusciel:2023umn, Harsh:2024kcl}, while Bondi--Sachs-type formulations have been developed to study radiation, asymptotic charges and mass loss beyond the linear regime \cite{Chrusciel:2016oux, Poole:2018koa, Compere:2019bua, Compere:2020lrt, Bonga:2023eml, McNees:2025acf, Poole:2025cmv}. A recurring theme  has been the search for a satisfactory notion of Bondi mass in the presence of a positive cosmological constant \cite{Szabados:2015wqa, Saw:2016isu, Saw:2017amv, Szabados:2018erf}. Geometric approaches based on the gravitational super-Poynting vector \cite{Fernandez-Alvarez:2019kdd, Fernandez-Alvarez:2020hsv, Wylleman:2020ubq, Fernandez-Alvarez:2021yog, Senovilla:2022pym, Fernandez-Alvarez:2023wal, Fernandez-Alvarez:2024bkf, Ciambelli:2024kre, Arenas-Henriquez:2025rpt, Fernandez-Alvarez:2025ivs} have discussed an independent characterisation of radiation in this setting. Significant attention has also been devoted to the asymptotic structure, boundary conditions, and symmetry algebras of asymptotically de Sitter spacetimes~\cite{Anninos:2011jp, Ashtekar:2014zfa,  PremaBalakrishnan:2019jvz, 
Compere:2019bua, Compere:2020lrt, Fernandez-Alvarez:2021uvz, Kaminski:2022tum, Bonga:2023eml, Hoque:2025iar}. Together, these developments have substantially clarified the conceptual and mathematical framework within which gravitational radiation and its associated charges are to be understood when $\Lambda > 0$. Nevertheless, the explicit multipolar characterisation of radiation from localised sources --- the de Sitter analogue of the multipolar hierarchy --- remains underdeveloped beyond quadrupolar order. The present paper is a step toward filling this gap.

It is fair to say that these complementary programmes have yet to converge on
a single characterisation of gravitational radiation. In asymptotically flat
general relativity, by contrast, one object --- the Bondi news --- plays
several roles at once. It provides an invariant criterion for the presence of
radiation, encodes the two radiative degrees of freedom, determines the loss
of Bondi mass, and drives the flux-balance laws of the BMS charges. For
$\Lambda > 0$, where conformal infinity is spacelike, it is not known whether
any single structure can play all of these roles.

The rest of the paper is organised as follows.  Since the octupolar analysis builds directly on the quadrupolar framework, we begin with a self-contained review of the quadrupolar setup in Section~\ref{sec:quadrupolar}. While the underlying calculations follow CHK, our presentation deliberately sets aside the discussion of residual gauge transformations and asymptotic symmetries, focusing instead on the construction of the metric perturbation in generalised harmonic gauge as a self-contained calculation. 
We hope this presentation serves as a useful entry point into the technical machinery of~\cite{Compere:2023ktn}. In Section~\ref{sec:SH_quad} we perform the spherical harmonic decomposition of the quadrupolar perturbation. In Section~\ref{sec:Octupolar} we derive the octupolar perturbation. In Section~\ref{sec:SH_oct} we perform its spherical harmonic decomposition. We conclude in Section~\ref{sec:conclusions}. Appendix~\ref{proof_identity} derives the universal multipole moment differentiation identity. Appendices~\ref{app:B}--\ref{app:D} develop the STF-to-spherical-harmonic translation technology. Appendix~\ref{STF-Mathematica} describes the Mathematica implementation of STF tensors used in our calculations.

\section{Quadrupolar radiation in de~Sitter spacetime}
\label{sec:quadrupolar}

This section reviews the derivation of the linearised gravitational waves 
around de~Sitter (dS) spacetime generated by a localised matter source, carried out at quadrupolar
order in the multipole expansion. We follow closely the treatment
of~\cite{Compere:2023ktn} and aim to present the material in a self-contained,
step-by-step manner. 

The structure of this section is as follows. Section~\ref{sec:perturb_eqs} formulates the perturbation equations in generalised harmonic gauge and identifies the three decoupled sectors. Section~\ref{sec:moments} defines the multipole moments of the stress-energy tensor, derives the flux-balance laws governing their time evolution, and establishes the consistent quadrupolar truncation. Sections~\ref{sec:scalar_vector} and~\ref{sec:tensor} derive the inhomogeneous solutions in the scalar/vector and tensor sectors, respectively. The extension of this framework to octupolar order is carried out in Section~\ref{sec:Octupolar}.

\subsection{Perturbation equations}
\label{sec:perturb_eqs}

We work exclusively in the future Poincar\'e patch of four-dimensional de~Sitter spacetime,
whose metric in conformally-flat  coordinates (also  called conformal de Sitter coordinates) $x^\mu=(\eta,x^i)$ reads
\begin{equation}
\bar{g}_{\alpha\beta}\,dx^\alpha dx^\beta
= a^2(\eta)\!\left(-d\eta^2 + d\vec{x}^2\right), \qquad
a(\eta) = -\frac{1}{H\eta},\qquad H=\sqrt{\frac{\Lambda}{3}},
\label{eq:dS_metric}
\end{equation}
where $\Lambda>0$ is the cosmological constant.  Conformal time runs over
$\eta\in(-\infty,0)$; future null infinity $\Scri$ is the spacelike surface $\eta\to 0$.  The radial coordinate is $\rho=\sqrt{x^i x^i}$.

Linear perturbations are written as $g_{\alpha\beta}=\bar{g}_{\alpha\beta}+h_{\alpha\beta}$.
Introducing the trace-reversed perturbation and the rescaled variable
\bea
\hat{h}_{\mu\nu} &:=& h_{\mu\nu}-\tfrac{1}{2}\bar{g}_{\mu\nu}(\bar{g}^{\alpha\beta}h_{\alpha\beta}), \\
\chi_{\mu\nu} &:=& a^{-2}\hat{h}_{\mu\nu},
\label{eq:chi_def}
\eea
the linearised Einstein equations simplify dramatically in the
generalised harmonic gauge \cite{deVega:1998ia}
\begin{equation}
\p^\alpha \chi_{\alpha\mu}
+ \frac{1}{\eta}\!\left(2\chi_{0\mu}+\delta^0_\mu\,\chi^\alpha{}_\alpha\right)=0.
\label{eq:gen_harmonic_gauge}
\end{equation}
In this gauge, the field equations decompose into three decoupled equations.
Define the combinations 
\be
\hat\chi:=\chi_{00}+ \delta^{ij} \chi_{ij},
\ee 
and $\hat{T}:=T_{00}+\delta^{ij}  T_{ij}$.
Then:
\begin{align}
\square\!\left(\frac{\hat\chi}{\eta}\right)
&= -\frac{16\pi G\,\hat{T}}{\eta},
\label{eq:scalar_eq}\\[4pt]
\square\!\left(\frac{\chi_{0i}}{\eta}\right)
&= -\frac{16\pi G\,T_{0i}}{\eta},
\label{eq:vector_eq}\\[4pt]
\left(\square+\frac{2}{\eta^2}\right)\frac{\chi_{ij}}{\eta}
&= -\frac{16\pi G}{\eta}\,T_{ij},
\label{eq:tensor_eq}
\end{align}
where $\square=-\p_\eta^2+\p_i^2$ is the flat Minkowski d'Alembertian.  Once
$\hat\chi$, $\chi_{0i}$ and $\chi_{ij}$ are known, the physical metric
perturbation is recovered from
\bea \label{eq:reconstruct_1}
h_{00}&=&\tfrac{1}{2}a^2\hat\chi, \\
h_{0i}&=&a^2\chi_{0i},\\
h_{ij}&=&a^2\chi_{ij}+a^2\delta_{ij}\!\left(\tfrac{1}{2}\hat\chi-\chi_{kk} \right).
\label{eq:reconstruct_3}
\eea

The key structural feature is that equations~\eqref{eq:scalar_eq} and \eqref{eq:vector_eq} are
formally identical to the Minkowski wave equation (with a modified source), while equation~\eqref{eq:tensor_eq} has an extra $2/\eta^2$ term that modifies the
Green's function and produces cosmological tail contributions. Throughout the paper, we raise and lower Cartesian indices with the three-dimensional flat metric $\delta_{ij}$.

\subsection{Multipole moments and conservation laws}
\label{sec:moments}

Because we are on a curved background, it is most natural to define source moments
using the orthonormal (tetrad) frame associated with the
Poincar\'e-patch metric \cite{Ashtekar:2015lxa, Date:2015kma}.  We define three families of multipole moments labelled by multi-index
$L=i_1 i_2\cdots i_l$ and $x_L=x_{i_1}\cdots x_{i_l}$:
\begin{align}
\Qrho_L
&:= \int d^3x\; a^{l+1}\, T_{00}\,x_L,
\label{eq:Qrho_def}\\
P_{i|L}
&:= \int d^3x\; a^{l+1}\, T_{0i}\,x_L,
\label{eq:P_def}\\
S_{ij|L}
&:= \int d^3x\; a^{l+1}\, T_{ij}\,x_L.
\label{eq:S_def}
\end{align}
The pressure moment 
\be
\Qp_L := S_{ii|L}=\delta^{ij}S_{ij|L},
\label{eq:Qrp_def}
\ee
and the combined moment \be 
\Qrp_L:=\Qrho_L+\Qp_L,
\ee
will appear frequently in our equations below.
For $l=0$: $\Qrho\equiv\Qrho_\emptyset$ is the total energy,
$\Qp\equiv\Qp_\emptyset$ the total pressure, and $P_i\equiv P_{i|\emptyset}$
the total momentum.

Covariant conservation $\nabla^\mu T_{\mu\nu}=0$ on the background~\eqref{eq:dS_metric}
reduces to the two partial-differential equations
\begin{align}
\p_\eta T_{00}-\p_i T_{0i}
-\frac{1}{\eta}\bigl(T_{00}+\delta^{ij}T_{ij}\bigr) &= 0,
\label{eq:Tcons1}\\
\p_\eta T_{0i}-\p_j T_{ij}-\frac{2}{\eta}T_{0i} &= 0.
\label{eq:Tcons2}
\end{align}
For many purposes, it is convenient to trade conformal time $\eta$ for cosmological (also called cosmic)  time 
$t$ via 
\be
\eta=-H^{-1}e^{-Ht},
\ee
so $\p_\eta = a(t)\,\p_t$.
Multiplying~\eqref{eq:Tcons1}--\eqref{eq:Tcons2} by $a^l x^L$
and integrating over $\vec{x}$, one obtains the flux-balance laws
for the multipole moments:
\begin{align}
\p_t \Qrho_L
&= H\!\left(l\,\Qrho_L - \Qp_L\right)
- l\, P_{(i_1|i_2\cdots i_l)},
\label{eq:Qrho_dot}\\
\p_t P_{i|L}
&= (l-1)H\, P_{i|L}
- l\, S_{i(i_1|i_2\cdots i_l)},
\label{eq:P_dot}
\end{align}
where round brackets denote symmetrisation and $S_{i\emptyset}:=0$ by
definition.

Solving~\eqref{eq:P_dot} for $S_{i(j|L)}$ we have:
\be
S_{i(j|L)} = -\frac{1}{l+1}(\p_t-l H)\,P_{i|jL},
\label{eq:S_asym}
\ee
and from   \eqref{eq:Qrho_dot} we have,
\be
S_{(ij|L)} = \frac{1}{(l+1)(l+2)}(\p_t-l H)\!
\left[(\p_t-(l+2)H)\Qrho_{ijL}+H\Qp_{ijL}\right].
\label{eq:S_sym}
\ee
These relations are indispensable for expressing the full solution
in terms of the irreducible set of moments.

The quadrupolar truncation is the assumption that all moments with
$l>2$ vanish~\cite{Compere:2023ktn}:
\begin{equation}
\int d^3x\; a^{l+1}\, T_{\mu\nu}\, x^L = 0,\qquad \forall \quad l>2.
\label{eq:quad_trunc}
\end{equation}
For an smooth extended source \eqref{eq:quad_trunc} is
 an approximation, and the neglected $l \ge 3$ moments enter the field at $\mathcal{O}(d^3)$, where $d$ is the bound on the coordinate size of the source at any retarded time.\footnote{The precise expression is below, cf.~equation \eqref{eq:omitted}.}  The truncation is consistent only if it is compatible with the
conservation equations, and the precise meaning of this requirement
is subtle. There has been some discussion of what is meant by
``consistent quadrupolar truncation''~\cite{Date:2024peu, Date:2026rft}.
The approach of CHK~\cite{Compere:2023ktn} is the cleanest in this regard, which is what we follow: in their
quadrupolar truncation, no spherical harmonics with $l>2$ appear in
the metric. This was already demonstrated in~\cite{Harsh:2024kcl},
albeit in a slightly different context, and will become clearer in
what follows.

We now collect several useful identities relating moments within the quadrupolar truncation.
Eq.~\eqref{eq:P_dot} for $l=1$ and $l=2$ respectively become,
\bea
\partial_t P_{i|j} &=& - S_{ij} \implies \partial_t P_{(i|j)} = - S_{ij}, \label{eq:P_dot_l=1}\\
\partial_t P_{i|jk} &=& H P_{i|jk} - S_{ij|k} - S_{ik|j}, \label{eq:P_dot_l=2}
\eea
and  Eq.~\eqref{eq:Qrho_dot} for  $l=2$ becomes,
\be \label{eq:Qrho_dot_l=2}
\partial_t Q_{ij}^{(\rho)} = - 2 P_{(i|j)} + H (2 Q_{ij}^{(\rho)} - Q_{ij}^{(p)}).
\ee
Together these two equations express $S_{ij}$ in terms of time derivatives of both the energy and pressure quadrupoles:
\begin{equation}
S_{ij} =  \frac{1}{2}\p_t\!\left(\p_t \Qrho_{ij}
- 2H\Qrho_{ij} + H\Qp_{ij}\right).
\label{eq:Sij_identity}
\end{equation}
This is a key equation we will use to write the tensor quadrupolar solution in Section \ref{sec:SH_quad}. 

Setting $l=3$ in the 
$\p_t \Qrho_L$ and $\partial_t P_{i|L}$ equations \eqref{eq:Qrho_dot} and \eqref{eq:P_dot} respectively, we get
\bea
\label{eq:cons_vanish_1}
&&    P_{(i|jk)} = 0, \\ 
&&   S_{i(j|kl)} = 0. 
\label{eq:cons_vanish_2}
\eea
From Eq.~\eqref{eq:P_dot_l=2} and
$P_{(i|jk)}=0$, one further deduces 
\be
S_{(ij|k)}=0. \label{eq:S-symmetric-zero}
\ee 
Equations~\eqref{eq:cons_vanish_1},  \eqref{eq:cons_vanish_2}, and~\eqref{eq:S-symmetric-zero} will
play an important role in the analysis that follows.

At this point Comp\`{e}re, Hoque, and Kutluk~\cite{Compere:2023ktn} decompose the quadrupolar moments into irreducible tensors in order to identify the minimal independent set. Rather than following their approach directly, we defer this analysis to Section~\ref{sec:SH_quad}, where we develop a slightly different approach to the same decomposition. The alternative is not only more transparent at quadrupolar order, but also is better suited for extension to octupolar truncation.

\subsection{Scalar and vector inhomogeneous solutions}
\label{sec:scalar_vector}

The scalar and vector field equations~\eqref{eq:scalar_eq}--\eqref{eq:vector_eq}
take the same form as the flat space wave equation with a $1/\eta$-modified
source.  Their retarded solutions are therefore given by the standard
Minkowski retarded Green's function (setting Newton's constant $G=1$ from now onwards):
\begin{align}
\hat\chi(\eta,\vec x)
&= 4\int\frac{d^3x'}{|\vec{x}-\vec{x}'|}
\frac{\eta}{\eta-|\vec{x}-\vec{x}'|}\,
\hat{T}\!\left(\eta-|\vec{x}-\vec{x}'|,\,\vec x'\right),
\label{eq:chi_hat_ret}\\
\chi_{0i}(\eta,\vec x)
&= 4\int\frac{d^3x'}{|\vec{x}-\vec{x}'|}
\frac{\eta}{\eta-|\vec{x}-\vec{x}'|}\,
T_{0i}\!\left(\eta-|\vec{x}-\vec{x}'|,\,\vec x'\right).
\label{eq:chi_0i_ret}
\end{align}
Let us define the retarded time $\eta_{\rm ret}:=\eta-\rho$ with $\rho=|\vec x|$. Let the coordinate size of the source  be bounded by  $d$ at any retarded time. The physical regime of interest is \cite{Ashtekar:2015lxa, Compere:2023ktn}:
\begin{enumerate}
\item The physical size
of the source is much smaller than the Hubble scale at any retarded time 
\be
a(\eta_{\rm ret})\,d\ll H^{-1}. \label{eq:266}
\ee
This implies $\frac{d}{-\eta_\ret} \ll 1$. \label{cond1}
\item \label{cond2} The observation point $\vec x$ is close to $\mathcal{I}^+$, i.e., 
$ \frac{-\eta}{\rho} \ll 1.$ This together with $\frac{d}{-\eta_\ret} \ll 1$ implies $d/\rho \ll 1$. 
\end{enumerate}
Under these conditions the integration in~\eqref{eq:chi_hat_ret}--\eqref{eq:chi_0i_ret}
can be expanded in powers of $d/\rho$ and $d/(-\eta_{\rm ret})$. Furthermore, the expansions can be 
consistently truncated at any given multipolar order.\footnote{More precisely, we now expand the integrand of~\eqref{eq:chi_hat_ret}--\eqref{eq:chi_0i_ret} in homogeneous polynomials
in $x'$ and truncate at the multipolar order of interest. At quadrupolar order we therefore
keep terms at most bilinear in $x'_i$.  Every term of degree $\ge 3$ in $x'$ integrates to
zero identically via \eqref{eq:quad_trunc}. Since 
the expansions terminate after finitely many terms and the question of convergence
does not arise.} At the quadrupolar order we only keep terms at most bilinear in
$x'_i$ and $\vec n\cdot \vec x'$, where $n_i=x_i/\rho$.  We introduce the shorthand
\begin{equation}
p := \vec{n} \cdot \vec{x}', \qquad
q := \rho'^2 - (\vec{n}\cdot\vec{x}')^2 = \rho'^2 - p^2 \;\geq\; 0,
\end{equation} 
where $\rho'^2 = \vec x' \cdot \vec x'$.  Note that $p \sim d$ and $q \sim d^2$. We have, 
\begin{align}
\label{eq:r_expand}
&  |\vec{x}-\vec{x}'| = \rho\!\left(
1
- \frac{p}{\rho}
+ \frac{q}{2\rho^2}
+ O\!\left(\frac{d^3}{\rho^3}\right)
\right), \\[4pt]
\label{eq:r_expand_inv}
& \frac{1}{|\vec{x}-\vec{x}'|}= \frac{1}{\rho}\!\left(
1
+ \frac{p}{\rho}
+ \frac{3p^2 - \rho'^2}{2\rho^2}
+ O\!\left(\frac{d^3}{\rho^3}\right)
\right),\\[4pt]
&  \frac{\eta}{\eta - |\vec{x}-\vec{x}'|} = \frac{\eta}{\eta_{\rm ret}}\!\left(
1
- \frac{p}{\eta_{\rm ret}}
+ \frac{p^2}{\eta_{\rm ret}^2}
+ \frac{q}{2\rho\,\eta_{\rm ret}}
+ \cdots 
\right),
\end{align}
where the omitted terms are schematically
\be
O\!\left(\frac{d^3}{\eta_{\rm ret}^3}, \; \frac{d}{\rho}\frac{d^2}{\eta_{\rm ret}^2},\;   \frac{d^2}{\rho^2}\frac{d}{\eta_{\rm ret}}\right). \label{eq:omitted}
\ee
The double expansion in $d/\rho$ and $d/(-\eta_{\rm ret})$ is what makes the de Sitter problem richer than the flat-space analogue. The omitted terms, schematically collected in \eqref{eq:omitted}, integrate to zero identically under the quadrupolar truncation \eqref{eq:quad_trunc}, so the truncated expressions below are exact rather than leading-order. However, in order to keep the presentation transparent we continue to write $\mathcal{O}(d^3)$ in several of the expressions below. The expressions are valid for $\rho > d \ge 0$.

Taylor expanding $\hat{T}$ around $\eta_\ret$ in the difference $\rho - | \vec x - \vec x'|$ (which is of the order of the source size $d$) to second order we have:
\begin{equation}\label{eq:Taylor_T}
\hat{T}(\eta - |\vec{x}-\vec{x}'|, \vec x')
= C_0 + C_1 + C_2 + O(d^3),
\end{equation}
where
\begin{align}
\label{eq:C0}C_0 &= \hat{T} (\eta_\ret, \vec x'), \\[4pt]
C_1 &= p\,\hat{T}^{(1)}(\eta_\ret, \vec x') ,
\label{eq:C1}\\[4pt]
C_2 &=  \frac{1}{2} \, p^2 \, \hat{T}^{(2)}(\eta_\ret, \vec x') -\frac{q}{2\rho}\,  \hat{T}^{(1)}(\eta_\ret, \vec x'),
\label{eq:C2}
\end{align}
and 
\be
\hat{T}^{(n)} = \partial_\eta^{n} \hat{T} \big{|}_{\eta = \eta_\ret}.
\ee

With these expansions, we write 
\be
\chihat = \frac{4\eta}{\rho \eta_\ret}\int d^3x'(G_0 + G_1 + G_2), 
\ee
truncated at quadrupolar order where $G_n$ collects all contributions of combined order $d^n$ in the source size $d$,
\begin{align}
\label{eq:G0}
G_0 &= \hat{T}^{(0)} \equiv \hat{T} (\eta_\ret, \vec x'), \\[4pt]
\label{eq:G1}
G_1 &= \left(\frac{p}{\rho} - \frac{p}{\eta_\ret}\right)\,\hat{T}^{(0)}
+ p\,\hat{T}^{(1)}, \\[4pt]
G_2 &= \left[\frac{3p^2-\rho'^2}{2\rho^2}
- \frac{3p^2-\rho'^2}{2\rho\,\eta_\ret}
+ \frac{p^2}{\eta_\ret^2}\right]\, \hat{T}^{(0)} 
+ \left[\frac{3p^2-\rho'^2}{2\rho}
- \frac{p^2}{\eta_\ret}\right]\, \hat{T}^{(1)}
+ \frac{1}{2} \, p^2 \, \hat{T}^{(2)}. \label{eq:G2}
\end{align}
In \cite{Compere:2023ktn}, the integrand $(G_0 + G_1 + G_2)$ is organised in inverse powers of $\rho$. For our purposes it is more natural to organise it instead by powers of the source size $d$, so that $G_n$ collects all contributions of combined order $d^n$. Under this scheme, extending the expansion to octupolar order requires only the addition of $G_3$, leaving $G_0$, $G_1$ and $G_2$ unchanged; each $G_n$ itself contains contributions at multiple powers of $\rho$.

We now perform the $\vec x'$ integrals and express $\hat\chi$ in terms of source moments. To this end, we require identities to convert $\eta$-derivatives of the source stress-energy tensor into source moments. From $\eta = -H^{-1}e^{-Ht}$ one has $\partial_\eta = a(t)\partial_t$.
The second $\eta$-derivative is therefore
\begin{equation}\label{eq:d2eta}
\partial^2_\eta
= \partial_\eta\bigl(a(t)\partial_t\bigr)
= a(t)\partial_t\bigl(a(t)\partial_t\bigr)
= a(t)\bigl[Ha(t)\partial_t + a(t)\partial^2_t\bigr]
= a^2(t)\bigl(\partial^2_t + H\partial_t\bigr).
\end{equation}
Such relations allow us to convert $\eta$-derivatives into $t$-derivatives. 

For the $t$-derivatives, we make use of the following identity:
\begin{equation}\label{eq:fundamental}
\int d^3x'\,a^{l+1}(t)\,x'^L\,\partial_t\hat{T}
= \partial_t \Qrp_{L} - (l+1)H\,\Qrp_{L}.
\end{equation}
This follows by differentiating the combined moment
\begin{equation}
\Qrp_{L}(t)= \int d^3x'\,a^{l+1}(t)\,x'^L\,\hat{T},
\end{equation}
with respect to $t$ and using $\partial_t a = Ha$, which gives
\begin{equation}
\partial_t \Qrp_{L}
= (l+1)H\,\Qrp_{L}
+ \int d^3x'\,a^{l+1}(t)\,x'^L\,\partial_t\hat{T}.
\end{equation}
Rearranging yields the identity~\eqref{eq:fundamental}.

Repeated application
of \eqref{eq:fundamental} allow us to write the integrals of $\hat T^{(n)}$ in terms of multipole moments and their $t$-derivatives. For example, applying \eqref{eq:fundamental} with $l = 1$ gives,
\begin{equation}
\int d^3x'\,a(\eta)\,x'_k\,\hat{T}^{(1)}
= \int d^3x'\,a^2(t)\,x'_k\,\partial_t\hat{T}
= \partial_t\Qrp_{k} - 2H\,\Qrp_{k}.
\end{equation}
Similarly, for the second $\eta$-derivative
\bea
\int d^3x'\,a(\eta)\,x'_k x'_l\, \hat{T}^{(2)} 
&=& \partial^2_t\Qrp_{kl} - 5H\,\partial_t\Qrp_{kl}
+ 6H^2\,\Qrp_{kl}.
\label{eq:273c}
\eea
A more general version of these identities is established in Appendix \ref{proof_identity}, and will be needed in later sections.

We introduce the shorthand $a_\ret := a(\eta_\ret)$ and define
\begin{equation}\label{eq:cR}
\frac{1}{r} := \frac{1}{a_\ret\rho} - H,
\end{equation}
which,
as will become clear when we pass to Bondi coordinates in Section~\ref{sec:SH_quad},  is the natural  radial variable to work with.
The overall prefactor in the integral for $\hat{\chi}$ simplifies as
\begin{equation}\label{eq:prefactor}
\frac{4\eta}{\rho\eta_\ret}
= 4a_\ret\!\left(\frac{1}{a_\ret\rho} - H\right)
= \frac{4a_\ret}{r}.
\end{equation}
Converting source integrals into moments, after some algebra we find
\begin{align}
\nonumber  \hat{\chi} =\;&
\frac{4}{r}\!\left[
\Qrp{}
- \frac{1}{2}H(\partial_t - H)\Qrp_{kk}
+ n_k\partial_t\Qrp_{k}
+ \frac{1}{2}n_kn_l(\partial^2_t - H^2)\Qrp_{kl}
\right]
\\[4pt]
&+  \frac{4}{r^2}\!\left[
n_k\Qrp_{k}
+ \frac{1}{2}(3n_kn_l-\delta_{kl})\partial_t\Qrp_{kl}
\right]+  \frac{2}{r^3}(3n_kn_l-\delta_{kl})\Qrp_{kl}, \label{chi-hat-quadrupolar}
\end{align}
where all moments are evaluated at $\eta_\ret$.

The derivation of \eqref{chi-hat-quadrupolar} for $\hat{\chi}$ carries over verbatim to
$\chi_{0i}$ under the replacement \be Q^{(\rho+p)}_L \to P_{i|L},\ee
since Eq.~\eqref{eq:chi_0i_ret} for $\chi_{0i}$ has the same structure as Eq.~\eqref{eq:chi_hat_ret} for $\hat{\chi}$ with \be 
\hat{T} \to T_{0i}. \ee 
The resulting
expression is simplified using the conservation equations \eqref{eq:P_dot_l=1}--\eqref{eq:P_dot_l=2} and the quadrupolar truncation condition $S_{(ij|k)} = 0$, cf.~Eq.~\eqref{eq:S-symmetric-zero}.  Using these substitutions we get our final form:
\begin{align}
\nonumber  \chi_{0i} =\;
&\frac{4}{r}\bigg[
P_i
- \frac{1}{2}H (\dt - H)P_{i|ll}
- n_kS_{ik}
+ \frac{1}{2}n_kn_l (\dt + H)S_{kl|i}
\bigg]\\[4pt]
+\;&\frac{4}{r^2}\bigg[
n_kP_{i|k}
- \frac{1}{2}\dt P_{i|ll}
+ \frac{3}{2}n_kn_l(S_{kl|i} + HP_{i|kl})
\bigg]
+\frac{2}{r^3}(3n^kn^l - \delta^{kl})P_{i|kl}, \label{eq:chi_0i_final}
\end{align}
where all moments are evaluated at $\eta_\ret$.

The scalar and vector perturbations Eq.~\eqref{chi-hat-quadrupolar} and Eq.~\eqref{eq:chi_0i_final} vanish on 
$\mathcal{I}^+$ (as $\eta \to 0$ or $r \to \infty$), so they leave no  imprint at 
future null infinity. This contrasts sharply with the tensor sector. It is this feature that makes the tensor inhomogeneous solution, to which 
we now turn, qualitatively richer than its scalar and vector counterparts.

\subsection{Tensor inhomogeneous solution}
\label{sec:tensor}
This section reviews the most involved part of the solution, namely the tensor component $\chi_{ij}$. The key qualitative difference from the scalar and vector sectors is the appearance of a cosmological ``tail'' sourced by the $+2/\eta^2$ term in the field equation~\eqref{eq:tensor_eq}. Rather than detailing the Green's function construction, we summarize its outcome (see, e.g., \cite{deVega:1998ia, Ashtekar:2015lxa, Compere:2023ktn} for derivations). The retarded solution naturally decomposes into two pieces: a sharp light cone contribution, $\chi^{(I)}_{ij}$, determined by the source evaluated at the retarded time, and a tail integral weighted by $1/\eta'^2$ over the interior of 
the past lightcone:
\be
4\int d^3x'\int_{-\infty}^{\eta- |\vec x - \vec x'|}
d\eta' \, \frac{T_{ij}(\eta', \vec x')}{\eta'^2} .
\ee
An integration by parts, valid under the assumption that $T_{ij}$ 
remains finite at the past cosmological horizon $\eta' \to -\infty$, converts the $\frac{1}{\eta'^2}$ weight into $\frac{1}{\eta'} \partial_{\eta'} $. The tail integral is then 
split  into a contribution 
running from $-\infty$ to $\eta_{\mathrm{ret}}$, which we denote $\chi^{(II)}_{ij}$, and a 
near-zone contribution running from $\eta_{\mathrm{ret}}$ to $\eta - |\vec{x} - \vec{x}'|$.  The latter is rendered more tractable by the shift $\eta' \mapsto \eta' + \eta_{\mathrm{ret}}$, 
producing $\chi^{(III)}_{ij}$ with its characteristic $\bigl(1 + \eta'/\eta_{\mathrm{ret}}\bigr)^{-1}$ 
kernel. Together, these three terms constitute the full particular solution \cite{Compere:2023ktn}
\begin{equation}
\chi_{ij} = \chi^{(I)}_{ij} + \chi^{(II)}_{ij} + \chi^{(III)}_{ij},
\label{eq:chi_split}
\end{equation}
where
\begin{align}
\chi^{(I)}_{ij}(\eta,\vec x)
&:= 4\int\frac{d^3x'}{|\vec x-\vec x'|}\,
T_{ij}\!\left(\eta-|\vec x-\vec x'|,\, \vec x'\right),
\label{eq:chiI_def}\\[4pt]
\chi^{(II)}_{ij}(\eta, \vec x)
&:= 4\int d^3x'\int_{-\infty}^{\eta_{\rm ret}}
\frac{d\eta'}{\eta'}\, \partial_{\eta'} T_{ij}(\eta',\vec x'),
\label{eq:chiII_def}\\[4pt]
\chi^{(III)}_{ij}(\eta, \vec x)
&:= \frac{4}{\eta_{\rm ret}}
\int d^3x'\int_0^{\rho-|\vec x-\vec x'|}
\frac{d\eta'}{1+\eta'/\eta_{\rm ret}}\,
\p_{\eta'} T_{ij}(\eta_{\rm ret}+\eta',\, \vec x').
\label{eq:chiIII_def}
\end{align}

\paragraph{The direct light cone term $\chi^{(I)}_{ij}$:}
Upon Taylor expanding  the retarded argument up to quadrupolar order we get,
\begin{equation}
T_{ij}(\eta - |\vec{x}-\vec{x}'|,\,\vec{x}')
= C_{0\:ij} + C_{1\:ij} + C_{2\:ij} + O(d^3),
\end{equation}
where
\begin{align}
C_{0\:ij} &= T^{(0)}_{ij}, &
C_{1\:ij} &= p\,T^{(1)}_{ij},&
C_{2\:ij} &= \frac{1}{2} \, p^2\,T^{(2)}_{ij} - \frac{q}{2\rho}\,T^{(1)}_{ij},
\end{align}
and $T^{(n)}_{ij} := \partial^{(n)}_\eta T_{ij}|_{\eta=\eta_{\rm ret}}$. The notation is chosen to parallel the $\hat T$ expansion discussed above.  Inserting the expansion of  $\frac{1}{|\vec x-\vec x'|}$  given in~\eqref{eq:r_expand_inv}, we write
\begin{equation}
\chi^{(I)}_{ij} = \frac{4}{\rho}\int d^3x'\,(G_{0\:ij} + G_{1\:ij} + G_{2\:ij}),
\end{equation}
where we employ the same $G_n$ organizational scheme introduced earlier, now applied to $T_{ij}$ in place of $\hat{T}$. The key structural difference is that, for $\chi^{(I)}_{ij}$, the integrand in~\eqref{eq:chiI_def} involves only the factor $1/|\vec{x}-\vec{x}'|$. Consequently, the $G_{n\,ij}$ blocks contain no contributions of the form $1/(\eta - |\vec{x}-\vec{x}'|)$, in contrast to $G_n$, cf.~\eqref{eq:chi_hat_ret}. We also note that the overall prefactor is now $4/\rho$. Explicitly, $G_{n\,ij}$ collects all contributions of combined order $d^n$ in the source size $d$,
\bea \label{eq:G0ij_1}
G_{0\:ij} &=& T^{(0)}_{ij},\\[6pt] \label{eq:G1ij_1}
G_{1\:ij} &=& \frac{p}{\rho}\,T^{(0)}_{ij}
+ p\,T^{(1)}_{ij},\\[6pt] \label{eq:G2ij_1}
G_{2\:ij} &=& \frac{3p^2-\rho'^2}{2\rho^2}\,T^{(0)}_{ij}
+ \frac{3p^2-\rho'^2}{2\rho}\,T^{(1)}_{ij}
+ \frac{1}{2} \, p^2\,T^{(2)}_{ij},
\eea
Extending to octupolar order requires only the addition of $G_{3\:ij}$,
leaving $G_{0\:ij}$, $G_{1\:ij}$ and $G_{2\:ij}$ unchanged.

Converting each integral to stress-energy moments via\footnote{A proof of this identity is given in Appendix \ref{proof_identity}.}
\begin{equation}\label{eq:identity}
\int d^3x'\,x'^L\,T^{(n)}_{ij} 
= a^{n-l-1}
\prod_{k=0}^{n-1}\!\bigl(\partial_t - (l+1-k)H\bigr)\,S_{ij|L} 
\end{equation}
and substituting $p = n_k x_k'$,
the integrals over $G_{0\:ij}$, $G_{1\:ij}$, $G_{2\:ij}$ yield
\bea \label{eq:G0ij}
\frac{4}{\rho}\int d^3x'\,G_{0 \: ij}
&=& \frac{4}{a_\ret \rho}\,S_{ij},\\[6pt]
\frac{4}{\rho}\int d^3x'\,G_{1\: ij} \label{eq:G1ij}
&=& \frac{4}{a_\ret \rho}\, n_k (\partial_t-2H)S_{ij|k}
+ \frac{4}{a_\ret^2\rho^2}\, n_k S_{ij|k},\\[6pt]
\frac{4}{\rho}\int d^3x'\,G_{2 \: ij}
&= & \frac{2}{a_\ret \rho}
n_k n_l (\partial^2_t-5H\partial_t+6H^2)\,S_{ij|kl}    + \frac{2}{a_\ret ^2\rho^2}
\, (3n_k n_l-\delta_{kl})     (\partial_t-3H)\,S_{ij|kl} \nonumber \\ 
&&    + \,  \frac{2}{a_\ret ^3\rho^3} \, (3n_k n_l-\delta_{kl}) S_{ij|kl},
\label{eq:G2ij}
\eea
where all terms are evaluated at $\eta_{\rm ret}$. Together these terms reproduce the result of~\cite{Compere:2023ktn}. 
The expressions are here written in terms of inverse powers of 
$a_{\rm ret}\rho$ rather than the radial variable 
$1/r = 1/(a_{\rm ret}\rho) - H$ defined in Eq.~\eqref{eq:cR}. The rewriting in terms of $1/r$ is deferred until all three 
contributions $\chi^{(I)}_{ij}$, $\chi^{(II)}_{ij}$ and 
$\chi^{(III)}_{ij}$ have been assembled, since extensive 
cancellations occur between them at that stage.

\paragraph{The global tail term $\chi^{(II)}_{ij}$:}
This is perhaps the most subtle term.  We first interchange the order of the $\vec x'$ and $\eta'$ integrations and  
apply the identity \eqref{eq:identity} with $l=0$ and $n=1$:
\bea
\int d^3x'\,\partial_\eta T_{ij}(\eta,x') &=&  (\partial_t - H) S_{ij} \\ 
&=&   \p_t \left[\frac{1}{2} \left(\partial_t - H\right) \!\left(\p_t \Qrho_{ij}
- 2H\Qrho_{ij} + H\Qp_{ij}\right) \right],
\label{eq:T_from_cons}
\eea
where in the last step we have used the identity \eqref{eq:Sij_identity} and commuted $\partial_t$ past $\partial_t - H$. 
Now we switch the $\eta$ integration to the $t$ integration. We note that $\frac{d\eta}{\eta} = - H dt$.  The integral then becomes a
total derivative (hence the term \emph{global tail}):
\bea
\label{eq:chiII_total_deriv}
\chi^{(II)}_{ij}
&=& -2H\int_{-\infty}^{t_{\rm ret}} dt\,\p_t\!\left[\left(\partial_t - H\right) \!\left(\p_t \Qrho_{ij}
- 2H\Qrho_{ij} + H\Qp_{ij}\right) \right], \\
&=&  -2H \left[\left(\partial_t - H\right) \!\left(\p_t \Qrho_{ij}
- 2H\Qrho_{ij} + H\Qp_{ij}\right) \right]_{\eta = \eta_\ret}  + \chi_{ij} (-\infty),
\eea
where
\be
\chi_{ij} (-\infty) = 2H \left[\left(\partial_t - H\right) \!\left(\p_t \Qrho_{ij}
- 2H\Qrho_{ij} + H\Qp_{ij}\right) \right]_{\eta = - \infty}.
\ee
This is a remarkable simplification. A seemingly complicated integral over
the entire past history of the source collapses to a boundary term.
Expanding out the derivative in \eqref{eq:chiII_total_deriv}, we get
\begin{equation}
\chi^{(II)}_{ij}
= -2H\!\left[\p_t^2\Qrho_{ij}
- 3H\p_t\Qrho_{ij}
+ 2H^2\Qrho_{ij}
+ H\p_t\Qp_{ij}
- H^2\Qp_{ij}\right]
+ \chi_{ij}(-\infty).
\label{eq:chiII_final}
\end{equation}
All terms are evaluated at $\eta_{\rm ret}$, except for $\chi_{ij} (-\infty)$, which  is the only non-instantaneous (i.e., non-local in retarded time)
contribution to the full solution.  It encodes the effect of the
source's entire past history extending back to the past cosmological horizon.

A point crucial for what follows is that the derivation of $\chi^{(II)}_{ij}$ is exact and does not rely on the quadrupolar truncation; it therefore carries over unchanged to octupolar order in Section~\ref{sec:Octupolar}.

\paragraph{The near-zone tail correction $\chi^{(III)}_{ij}$:}
This term captures the tail correction from the near zone, in the sense 
that the $\eta'$ integral runs only over the interval $[0,\delta]$, where
\begin{equation} \label{eq:delta}
\delta := \rho - |\vec{x} - \vec{x}'|.
\end{equation}
From Eq.~\eqref{eq:r_expand} we already know that
\begin{equation}
\delta = p - \frac{q}{2\rho} + O(d^3),
\end{equation}
so that $\delta = O(d)$ is small in the source-size expansion. Since the 
integration range is itself of order $d$, it suffices to expand
\begin{equation}
\partial_{\eta'} T_{ij}(\eta_{\rm ret}+\eta',\,\vec{x}')
\end{equation}
in a Taylor series around $\eta' = 0$, retaining only as many terms as 
are needed at the desired multipolar order. At quadrupolar order, 
truncation at first order in $\eta'$ is sufficient, giving
\begin{equation}
\partial_{\eta'} T_{ij}(\eta_{\rm ret}+\eta',\,\vec{x}') 
= T^{(1)}_{ij} + \eta'\,T^{(2)}_{ij} + O(\eta'^2).
\end{equation}
The full expansion of the integrand in \eqref{eq:chiIII_def} reads 
\be \label{eq:integrand_expansion}
\frac{1}{1 + \eta'/\eta_{\ret}} \p_{\eta'} T_{ij}(\eta_{\rm ret}+\eta',\, \vec x')  = T^{(1)}_{ij} + \eta'\left(T^{(2)}_{ij} - \frac{1}{\eta_\ret} T^{(1)}_{ij}\right) + O(\eta'^2).
\ee
The final integration over $\eta' \in [0,\delta]$ is straightforward and yields,
\be
\int_0^\delta d \eta' \frac{1}{1 + \eta'/\eta_{\ret}} \p_{\eta'} T_{ij}(\eta_{\rm ret}+\eta',\, \vec x')  = T^{(1)}_{ij} \delta +\frac{1}{2} \left(T^{(2)}_{ij} - \frac{1}{\eta_\ret} T^{(1)}_{ij}\right) \delta^2 + \cdots.
\ee
Substituting for $\delta$ and using the moment identity \eqref{eq:identity} repeatedly gives,
\begin{align}
\chi^{(III)}_{ij}
= & -4H\bigg[
n_k(\p_t -2H)S_{ij|k}
- \frac{1}{2a_\ret \rho}(\delta_{kl}-n_k n_l)(\p_t -3H) S_{ij|kl}\nonumber \\
& + \frac{1}{2}n_k n_l(\p_t^2-4H\p_t+3H^2)S_{ij|kl}
\bigg].
\label{eq:chiIII_final}
\end{align}

Adding all three parts, one observes extensive cancellations. Before writing the final combined result, it is useful to isolate the 
part of the tensor perturbation that survives on $\mathcal{I}^+$. 
Inspection of the three contributions shows that all terms carrying 
positive powers of $1/r$ vanish as $r \to \infty$, 
while the part independent of 
$1/r$ remains finite and encodes the 
imprint of radiation at future null infinity. We therefore define the 
effective boundary datum
\begin{equation}
\chi^{(0)}_{ij}
:= \chi_{ij}(-\infty)
+ 4H^2\!\left[
-P_{(i|j)}
+ n_k S_{ij|k}
+ \frac{1}{2}n_k n_l\p_t S_{ij|kl}
- \frac{1}{2} H S_{ij|kk}
\right]_{\eta_{\rm ret}}.
\label{eq:chi0_def}
\end{equation}
This quantity is $O(H^2)$ and vanishes in the flat limit. It 
satisfies the conservation law
\begin{equation}
\p_t\chi^{(0)}_{ij}
= 4H^2\!\left[S_{ij}
+ n_k\p_t S_{ij|k}
+ \frac{1}{2}n_k n_l \p_t^2 S_{ij|kl}
- \frac{1}{2} H\p_t S_{ij|kk}\right].
\label{eq:chi0_cons}
\end{equation}
The complete tensor solution then takes the compact form:
\bea
\chi_{ij} &=& \chi^{(0)}_{ij}
+ \frac{1}{r}H^{-2}\p_t\chi^{(0)}_{ij}
+ \frac{4}{r^2}
\left[n_k S_{ij|k}+\frac{1}{2}(3n_k n_l-\delta_{kl})\p_t S_{ij|kl}\right] \nonumber \\
&& + \frac{2}{r^3}
(3n_k n_l-\delta_{kl})\,S_{ij|kl}.
\label{eq:chi_ij_final}
\eea
Equation~\eqref{eq:chi_ij_final} is exact to all orders in $1/r$
within the quadrupolar approximation.

This completes the derivation of the linearised metric perturbation at quadrupolar order in generalised harmonic gauge. The solution is expressed in STF tensor language, which is natural from the perspective of the field equations but not immediately suited for comparison with the broader de Sitter literature or for rendering the angular structure of the radiation manifest. Section \ref{sec:SH_quad} develops the translation into spherical harmonic form.

\section{Spherical harmonic decomposition of the quadrupolar perturbation}
\label{sec:SH_quad}
The solutions for $\hat\chi$, $\chi_{0i}$, and $\chi_{ij}$ ---
Eqs.~\eqref{chi-hat-quadrupolar}, \eqref{eq:chi_0i_final}, and
\eqref{eq:chi_ij_final} respectively --- are expressed in terms of Cartesian
unit vector components $n^k = x^k/\rho$ contracted with multipole moments. For
practical computations and for comparison with the wider de~Sitter
gravitational wave literature, in particular the Bondi--Sachs formalism, it is
useful to express these angular structures in spherical polar coordinates
$(\rho,\theta,\phi)$, where
\begin{equation}
  \vec{x} = (\rho\sin\theta\cos\phi,\,\rho\sin\theta\sin\phi,\,\rho\cos\theta).
\end{equation}
Furthermore, since the Bondi form is the natural setting for analysing
radiation at $\mathcal{I}^+$, we also write the background de~Sitter metric in
Bondi form. The coordinate transformation from conformal coordinates to Bondi
coordinates is
\begin{align}\label{eq:transformation}
  \eta &= -\frac{e^{-Hu}}{H(1+Hr)},&  \rho &= \frac{re^{-Hu}}{1+Hr},
\end{align}
with angles $x^A = (\theta,\phi)$ unchanged. Note that
$\eta_{\rm ret}=\eta-\rho=-e^{-Hu}/H$, so that $u$ is the retarded time and
$a_{\rm ret}=e^{Hu}$. Moreover, one readily checks that
$1/(a_{\rm ret}\rho)-H=1/r$: the Bondi radial coordinate coincides with the
radial variable introduced in Eq.~\eqref{eq:cR}, which is why the latter was
designated the natural radial variable. The de~Sitter metric takes the form
\begin{equation}
  d\bar{s}^2 := (H^2r^2-1)du^2 - 2\,du\,dr 
  + r^2\mathring{q}_{AB}dx^A dx^B,
\end{equation}
where $\mathring{q}_{AB}dx^A dx^B = d\theta^2 + \sin^2\theta\,d\phi^2$ 
is the round metric on the two-sphere. We introduce $\perp^{ij} := 
\delta^{ij} - n^i n^j$ and $e^i_A := \partial n^i/\partial x^A$, which 
satisfies $e^i_A n_i = 0$. The Jacobian of the transformation \eqref{eq:transformation} is
\begin{align}
  \frac{\partial\eta}{\partial u} &= \frac{e^{-Hu}}{1+Hr}, &
  \frac{\partial\eta}{\partial r} &= \frac{e^{-Hu}}{(1+Hr)^2},\\[4pt]
  \frac{\partial\rho}{\partial u} &= \frac{-Hre^{-Hu}}{1+Hr}, &
  \frac{\partial\rho}{\partial r} &= \frac{e^{-Hu}}{(1+Hr)^2}.
\end{align}
Substituting the Jacobian and using \eqref{eq:reconstruct_1}--\eqref{eq:reconstruct_3},  
one finds the metric perturbation components in $(u,r,\theta,\phi)$ 
coordinates:
\begin{align}
\label{huu}
 h_{uu} &= \frac{1}{2}(1+H^2r^2)\,\hat{\chi}
           - 2Hr\,n^i\chi_{0i} 
           - H^2r^2\perp^{ij}\chi_{ij},\\[6pt]
 h_{ur} &= \frac{1}{2}\left(\frac{1-Hr}{1+Hr}\right)
            (\hat{\chi} + 2n^i\chi_{0i}) 
           + \frac{Hr}{1+Hr}\perp^{ij}\chi_{ij},\\[6pt]
 h_{rr} &= \frac{1}{(1+Hr)^2}
            \left[\hat{\chi} + 2n^i\chi_{0i} 
            - \perp^{ij}\chi_{ij}\right],\\[6pt]
\label{huA}
 h_{uA} &= re^i_A\chi_{0i} - Hr^2n^ie^j_A\chi_{ij},\\[6pt]
\label{hrA}
 h_{rA} &= \frac{r}{1+Hr}\,e^i_A(\chi_{0i} + n_j\chi_{ij}),\\[6pt]
 h_{AB} &= r^2(e^i_{(A}e^j_{B)} - \mathring{q}_{AB}\delta_{ij})\chi_{ij}
           + \frac{1}{2}r^2\mathring{q}_{AB}\hat{\chi}.
\label{hAB}
\end{align}

This form of the perturbation is well suited for further analysis. In what
follows, we express the right-hand sides of Eqs.~\eqref{huu}--\eqref{hAB} in
terms of spherical harmonics. Under the quadrupolar truncation, only spherical
harmonics (scalar, vector, and tensor) with $l\le 2$ contribute. This feature
was already demonstrated in~\cite{Harsh:2024kcl}, albeit in a different gauge
and context. We emphasise that \emph{the metric perturbation
\eqref{huu}--\eqref{hAB} is not in Bondi gauge; only the background de~Sitter
metric is in Bondi gauge}. Converting the metric perturbation to Bondi gauge is
a separate exercise \cite{Compere:2023ktn}, which we do not pursue here. Since
the background is in Bondi gauge, we nevertheless refer to $(u,r,\theta,\phi)$
as Bondi coordinates.

The projections that appear repeatedly in the Bondi components are
\begin{equation}
  \hat{\chi},\quad n^i\chi_{0i},\quad e^i_A\chi_{0i},\quad
  \delta^{ij}\chi_{ij},\quad n^i n^j\chi_{ij},\quad
  n^i e^j_A\chi_{ij},\quad e^{i}_{\langle A}e^{j}_{B\rangle}\,
             \chi_{ij}.
  \label{eq:quad-projection-building-blocks}
\end{equation}
To apply the STF-to-spherical-harmonic identities of
Appendices~\ref{app:B}--\ref{app:D}, it is useful to separate the factors that
depend on $r$ and $t$ from the angular STF factors. In the following forms,
all moments are evaluated at retarded time, and
$\widehat n_{ij} := n_i n_j - \tfrac{1}{3}\delta_{ij}$ denotes the
symmetric trace-free part of $n_i n_j$ (the general STF product
$\widehat n_L$ is discussed in Appendix~\ref{app:B}). At quadrupolar order the
scalar solution can be written as
\begingroup\small
\begin{align}
  \hat\chi
  &= \frac{4}{r}\Qrp{}
   + \frac{4}{r}\left(\partial_u+\frac{1}{r}\right)\left(n_i\Qrp_i\right)
   + \frac{2}{r}\left(\partial_u^2+\frac{3}{r}\partial_u+\frac{3}{r^2}-H^2\right)
     \left(\widehat n_{ij}\Qrp_{ij}\right) \nonumber\\
  &\quad
   + \frac{2}{3r}(\partial_u-H)(\partial_u-2H)
     \left(\delta_{ij}\Qrp_{ij}\right).
  \label{eq:sec3-chihat-quad-operator-form}
\end{align}
\endgroup
The vector solution takes the same
form, with the replacement $\Qrp_L\to P_{i|L}$:
\begingroup\small
\begin{align}
  \chi_{0i}
  &= \frac{4}{r}P_i
   + \frac{4}{r}\left(\partial_u+\frac{1}{r}\right)\left(n_jP_{i|j}\right)
   + \frac{2}{r}\left(\partial_u^2+\frac{3}{r}\partial_u+\frac{3}{r^2}-H^2\right)
     \left(\widehat n_{jk}P_{i|jk}\right) \nonumber\\
  &\quad
   + \frac{2}{3r}(\partial_u-H)(\partial_u-2H)
     \left(\delta_{jk}P_{i|jk}\right).
  \label{eq:sec3-chi0i-quad-operator-form}
\end{align}
\endgroup
For the tensor solution we keep the symmetric moment \(P_{(i|j)}\) explicit
via $S_{ij} = - \partial_u P_{(i|j)}$:
\begingroup\small
\begin{align}
  \chi_{ij} ={}& \left(-4H^2-\frac{4}{r}\partial_u\right)P_{(i|j)}
    +4\left(\frac{1}{r}\partial_u+\frac{1}{r^2}+H^2\right)
      \left(n_kS_{ij|k}\right) \nonumber\\
    &+\frac{2}{3}\left(\frac{1}{r}\partial_u+H^2\right)(\partial_u-3H)
      \left(\delta_{kl}S_{ij|kl}\right) \nonumber\\
    &+2\left[
      \frac{1}{r}\partial_u^2+\left(H^2+\frac{3}{r^2}\right)\partial_u
      +\frac{3}{r^3}
      \right]\left(\widehat n_{kl}S_{ij|kl}\right).
  \label{eq:sec3-chiij-quad-operator-form}
\end{align}
\endgroup

This form makes the subsequent projections transparent. The differential
factors depend only on $r$ and retarded time, while the STF factors carry the
spherical-harmonic content. Here and in what follows we set
$\chi_{ij}(-\infty)=0$, as appropriate for a source whose multipole moments
and their time derivatives vanish in the asymptotic past. More generally,
this constant term can be reinstated additively without affecting the
analysis below.

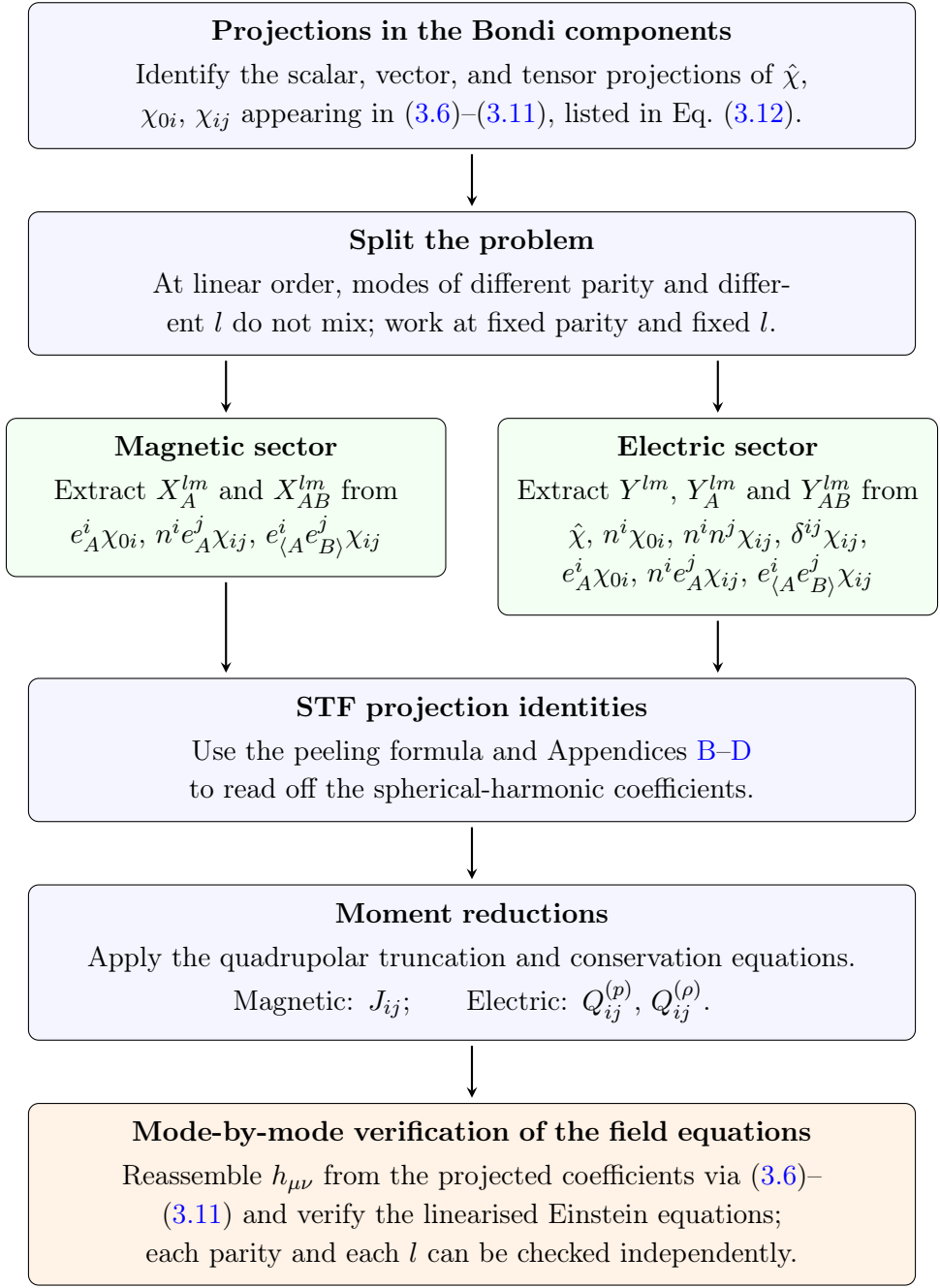
\begin{figure}[t]
\centering
\begin{tikzpicture}[
  >=stealth, node distance=0.85cm,
  flow/.style={draw, rounded corners, align=center, fill=blue!4,
    text width=11.6cm, inner sep=7pt},
  branch/.style={draw, rounded corners, align=center, fill=green!5,
    text width=5.5cm, inner sep=7pt},
  result/.style={draw, rounded corners, align=center, fill=orange!10,
    text width=11.6cm, inner sep=7pt},
  arrow/.style={->, thick, shorten <=2pt, shorten >=2pt}
]
  \node[flow] (bondi)
  {\textbf{Projections in the Bondi components}\\[2pt]
  Identify the scalar, vector, and tensor projections of $\hat\chi$,
  $\chi_{0i}$, $\chi_{ij}$ appearing in \eqref{huu}--\eqref{hAB},
  listed in Eq.~\eqref{eq:quad-projection-building-blocks}.};

  \node[flow, below=of bondi] (split)
  {\textbf{Split the problem}\\[2pt]
  At linear order, modes of different parity and different $l$ do not mix;
  work at fixed parity and fixed $l$.};

  \node[branch, anchor=north east] (mag)
    at ([xshift=-0.35cm, yshift=-0.85cm]split.south)
  {\textbf{Magnetic sector}\\[2pt]
  Extract $X_A^{lm}$ and $X_{AB}^{lm}$ from\\[2pt]
  $e_A^i\chi_{0i}$, $n^ie_A^j\chi_{ij}$,
  $e_{\langle A}^ie_{B\rangle}^j\chi_{ij}$};

  \node[branch, anchor=north west] (elec)
    at ([xshift=0.35cm, yshift=-0.85cm]split.south)
  {\textbf{Electric sector}\\[2pt]
  Extract $Y^{lm}$, $Y_A^{lm}$ and $Y_{AB}^{lm}$ from\\[2pt]
  $\hat\chi$, $n^i\chi_{0i}$, $n^in^j\chi_{ij}$,
  $\delta^{ij}\chi_{ij}$, $e_A^i\chi_{0i}$,
  $n^ie_A^j\chi_{ij}$, $e_{\langle A}^ie_{B\rangle}^j\chi_{ij}$};

  \node[flow, anchor=north] (project)
    at ([yshift=-0.85cm]elec.south -| split)
  {\textbf{STF projection identities}\\[2pt]
  Use the peeling formula and Appendices~\ref{app:B}--\ref{app:D} to read off
  the spherical-harmonic coefficients.};

  \node[flow, below=of project] (reduce)
  {\textbf{Moment reductions}\\[2pt]
  Apply the quadrupolar truncation and conservation equations.\\[2pt]
  Magnetic: $J_{ij}$;\qquad Electric:
  $Q^{(p)}_{ij}$, $Q^{(\rho)}_{ij}$.};

  \node[result, below=of reduce] (assemble)
  {\textbf{Mode-by-mode verification of the field equations}\\[2pt]
  Reassemble $h_{\mu\nu}$ from the projected coefficients via
  \eqref{huu}--\eqref{hAB} and verify the linearised Einstein equations;
  each parity and each $l$ can be checked independently.};

  \draw[arrow] (bondi) -- (split);
  \draw[arrow] (split.south -| mag) -- (mag.north);
  \draw[arrow] (split.south -| elec) -- (elec.north);
  \draw[arrow] (mag.south) -- (mag.south |- project.north);
  \draw[arrow] (elec.south) -- (elec.south |- project.north);
  \draw[arrow] (project) -- (reduce);
  \draw[arrow] (reduce) -- (assemble);
\end{tikzpicture}
\caption{Flow chart of the quadrupolar spherical-harmonic decomposition
carried out in this section.}
\label{fig:quad-sh-flow-chart}
\end{figure}

The remainder of this section carries out the decomposition summarised in
Figure~\ref{fig:quad-sh-flow-chart}. Starting from the projections
\eqref{eq:quad-projection-building-blocks} of the expressions~\eqref{eq:sec3-chihat-quad-operator-form}--\eqref{eq:sec3-chiij-quad-operator-form},
we work at fixed parity and fixed $l$. The magnetic   sector is
built on the vector and tensor harmonics $X^{lm}_A$ and $X^{lm}_{AB}$, and the
electric   sector on $Y^{lm}$, $Y^{lm}_A$ and $Y^{lm}_{AB}$. The
spherical-harmonic coefficients are extracted using the peeling formula and
the STF projection identities of Appendices~\ref{app:B}--\ref{app:D}, and are
then simplified using the quadrupolar truncation and the conservation
equations. As a check on the entire construction, we
 verify the linearised Einstein equations mode by
mode. The same strategy, with octupolar moments, is followed in
Section~\ref{sec:SH_oct}.

\subsection{Magnetic sector}
We begin with the magnetic sector, treating the $l=2$ and $l=1$ modes in
turn. There is no $l=0$ magnetic mode. The analysis relies on the identities
of Appendices~\ref{app:B}--\ref{app:D}, which decompose contractions
involving $n^i$ and $e^j_A$ into irreducible tensor harmonics.
\subsubsection{\texorpdfstring{$l=2$}{l=2} terms}

The central tool for extracting the spherical-harmonic content is identity~\eqref{final_C19}. This
identity decomposes the contraction of a sphere tangent vector $e^i_A$ and the
STF unit vector product $\widehat{n}^L$ with an arbitrary rank-$(l+1)$ tensor
$P_{iL}$ into three irreducible pieces, corresponding to harmonics of order
$l+1$, $l$, and $l-1$ respectively. The expressions for $h_{uA}$, $h_{rA}$,
and $h_{AB}$ then indicate that extracting the $l=2$ magnetic-parity sector
requires analysing only the following three structures:
\begin{align}
& e^{i}_{A}\,\chi_{0i}, &  & n^{i}e^{j}_{A}\,\chi_{ij} ,  & & e^{i}_{\langle A}e^{j}_{B\rangle}\,
             \chi_{ij}. &
\end{align}

We begin with $e^{i}_{A}\,\chi_{0i}$. Of the four terms in
\eqref{eq:sec3-chi0i-quad-operator-form}, only the STF two-unit-vector
contribution carries the structure that identity~\eqref{final_C19} maps to the
$l=2$ magnetic harmonic $X^{l=2}_A$. This part is
\begin{equation}
  \chi_{0i}\Big{|}_{\widehat n_{kl}}
  = \frac{2}{r}\left(\partial_u^2+\frac{3}{r}\partial_u+\frac{3}{r^2}-H^2\right)
    \left(\widehat n_{kl}P_{i|kl}\right).
\end{equation}
For an arbitrary tensor $T_{ikl}$, identity \eqref{final_C19} gives the
coefficient of $X_{A}^{\langle ij \rangle}$ in the contraction
$e^{i}_{A}\,\widehat n^{kl}\,T_{ikl}$ as
\begin{equation}
 e^{i}_{A}\,\widehat n^{kl}\,T_{ikl}\Big{|}_{X^{l=2}_{A}}
  = -\frac{1}{3}\,\STF_{ij}\!\left\{\epsilon_{ikl}\,T_{k\langle jl \rangle}
    \right\}X_{A}^{\langle ij \rangle}\,.
  \label{eq:313}
\end{equation}
When $T_{ikl} = P_{i|kl}$, which is symmetric in
$(k,l)$, we have simply
\begin{equation}
 e^{i}_{A}\,\widehat n^{kl}\,P_{i|kl}\Big{|}_{X^{l=2}_{A}}
  = -\frac{1}{3}\,\STF_{ij}\!\left\{\epsilon_{ikl}\,P_{k|lj}
    \right\}X_{A}^{\langle ij \rangle}\, \equiv - \frac{1}{4} J_{ij}  X_{A}^{\langle ij \rangle},
  \label{eq:314}
\end{equation}
where, following CHK, we have defined the current quadrupole
\be
J_{ij} = \frac{4}{3} \STF_{ij} \left[ \epsilon_{ikl} P_{k|lj}\right],
\ee
retaining the factor of $4/3$ of the CHK convention for ease of comparison.
The key point is that there is no need to separately analyse the irreducible
decomposition of $P_{i|jk}$ and the other tensors. Instead, the
spherical-harmonic coefficients naturally project onto the relevant
irreducible components. This differs slightly from the strategy adopted by
CHK~\cite{Compere:2023ktn}, where the tensorial decomposition is performed
explicitly. Although the two procedures are equivalent, projecting onto the
relevant irreducible component is easier, and it extends directly to the
octupolar truncation considered later. Since $\partial_t$ commutes with the
$\STF$ projection, applying \eqref{eq:314} directly
yields the first structure:
\begin{equation}
  e^{i}_{A}\,\chi_{0i}\Big{|}_{X^{l=2}_{A}}
  = - \frac{1}{2r}\left[(\partial_{u}^{2}-H^{2})
          +\frac{3}{r} \partial_{u}
          +\frac{3}{r^{2}}
    \right]\,J_{ij}\,X_A^{\langle ij \rangle}.
\end{equation}

For the remaining structures it is convenient to introduce, again following
CHK,
\be
K_{ij} = \frac{4}{3} \STF_{ij} \left[ \epsilon_{ikl} S_{jk|l}\right].
\ee
From the conservation equation~\eqref{eq:P_dot_l=2}, together with the symmetry of $S_{ij|k}$ in its
first two indices, it follows that
\begin{equation}
(\partial_u - H) J_{ij} = -K_{ij},
\label{eq:JKrelation}
\end{equation}
so that all $l=2$ magnetic coefficients can be expressed through $J_{ij}$
alone.

For $e_A^i n^j\chi_{ij}$, the monopole and trace parts of $\chi_{ij}$ are
independent of $n^i$ and cannot produce an $l=2$ harmonic under contraction
with $ e^i_A n^j$. The relevant part is therefore the one-vector part,
\begin{equation}
  \chi_{ij}\Big{|}_{n_k}
  = 4\left(\frac{1}{r}\partial_u+\frac{1}{r^2}+H^2\right)
    \left(n_kS_{ij|k}\right).
\end{equation}
The contraction with $n^jn^k$ symmetrises $S_{ij|k}$ over the pair $(j,k)$,
and by Eq.~\eqref{eq:S_asym},
\be
S_{i(j|k)} = -\frac{1}{2}(\partial_u - H)\,P_{i|jk}.
\ee
Applying
\eqref{eq:314} then gives
\begin{equation}
  e^{i}_{A} n^j \chi_{ij}\Big{|}_{X^{l=2}_{A}}
  = \frac{1}{2r^2}(\partial_u - H) \!\left[H^{2}r^{2}+r\,\partial_{t}+1\right]J_{ij}X_A^{\langle ij \rangle}.
\end{equation}

The analysis of $e^{i}_{\langle A}e^{j}_{B\rangle}\,\chi_{ij}$ uses
Appendix~\ref{app:D}. In particular, Eq.~\eqref{final_D30} decomposes
combinations of the form $e^i_A e^j_B \widehat n_L$ into irreducible tensor
harmonics. As with the vector structure, only the one-vector part of
$\chi_{ij}$ contributes, and identity \eqref{final_D30} yields
\be
e^{i}_{\langle A}e^{j}_{B\rangle}\,
             \chi_{ij} \Big{|}_{X^{l=2}_{AB}} = \frac{1}{r^2}(\partial_u - H) \!\left[H^{2}r^{2}+r\,\partial_{u}+1\right]J_{ij}X_{AB}^{\langle ij \rangle}.
\label{eq323}
\ee

Substituting these expressions into $h_{uA}$, $h_{rA}$, and $h_{AB}$ yields
the purely $l=2$ contributions in the magnetic sector. Direct substitution
verifies that these terms satisfy the linearised Einstein equations. This
provides a nontrivial consistency check of the framework developed in this
paper.

\subsubsection{\texorpdfstring{$l=1$}{l=1} terms}
The $l=1$ magnetic sector is simpler. There are no
magnetic tensor harmonics at $l=1$, so $h_{AB}$ receives no $l=1$ magnetic
contribution, and the entire sector resides in $h_{uA}$ and $h_{rA}$ through
the two vector structures $e^i_A\chi_{0i}$ and $n^i e^j_A\chi_{ij}$.

Before evaluating these structures, we record the conservation law that
governs the sector. The $l=1$ case of the momentum equation~\eqref{eq:P_dot}
is
\begin{equation}
\partial_u P_{i|j}=-S_{ij}.
\label{eq:l1_Pdot}
\end{equation}
Since $S_{ij}$ is symmetric, the antisymmetric part of $P_{i|j}$ is conserved,
$\partial_u P_{[i|j]}=0$. Following~\cite{Compere:2023ktn, Harsh:2024kcl}, we
introduce the odd-parity (magnetic) dipole moment
\begin{equation}
J_i := \epsilon_{ijk}\,P_{j|k},
\label{eq:angular_momentum}
\end{equation}
which encodes the angular momentum of the source and, by~\eqref{eq:l1_Pdot},
is conserved:
\begin{equation}
\partial_u J_i = \epsilon_{ijk}\,\partial_u P_{j|k}= 0.
\label{eq:Ji_conserved}
\end{equation}
The $l=1$ magnetic perturbation is therefore stationary.

The relevant projection is the $l=1$ specialisation of
identity~\eqref{final_C19}. For an arbitrary rank-2 Cartesian tensor $T_{ab}$,
it gives the coefficient of the magnetic vector harmonic
$X^{\langle i\rangle}_A$ in the contraction $e^a_A\,n^b\,T_{ab}$ as
\begin{equation}
e^a_A\,n^b\,T_{ab}\,\big|_{X^{l=1}_A}
  = -\frac{1}{2}\,\epsilon_{iab}\,T_{ab}\,X^{\langle i\rangle}_A.
\label{eq:l1_mag_projection}
\end{equation}
The one-vector contribution in $\chi_{0i}$ is
\begin{equation}
  \chi_{0i}\Big{|}_{n_j}
  = \frac{4}{r}\left(\partial_u+\frac{1}{r}\right)
    \left(n_jP_{i|j}\right).
\end{equation}
Only the antisymmetric part of $P_{i|j}$ survives the contraction
with~\eqref{eq:l1_mag_projection}, so this structure is governed by $J_i$:
\begin{equation}
e^i_A\,\chi_{0i}\,\big|_{X^{l=1}_A}
  = -\frac{2}{r}\left(\partial_u+\frac{1}{r}\right)J_i\,X^{\langle i\rangle}_A
  = -\frac{2}{r^2}\,J_i\,X^{\langle i\rangle}_A,
\label{eq:l1_chi0i}
\end{equation}
where the second equality uses the conservation
law~\eqref{eq:Ji_conserved}.

The companion structure vanishes,
\begin{equation}
  n^i e^j_A \chi_{ij} \Big|_{X^{l=1}_A}=0,
\label{eq:companion-l1}
\end{equation}
although the cancellation is less immediate than it appears. In the 
form~\eqref{eq:sec3-chiij-quad-operator-form}, the $P_{(i|j)}$ and trace terms
enter the $l=1$ magnetic projection only through symmetric rank-two tensors
and are annihilated by~\eqref{eq:l1_mag_projection}. For the remaining term,
proportional to $\widehat n_{kl}S_{ij|kl}$, the contraction with $n^i e^j_A$
must be reduced to its rank-one angular part, which yields a contribution
proportional to $e^j_A n^l\big(S_{aj|al}-\tfrac{1}{3}S_{jl|aa}\big)$. The
trace term $S_{jl|aa}$ is symmetric in $(j,l)$ and hence annihilated
by~\eqref{eq:l1_mag_projection}. The symmetry of the first term is not
guaranteed by the moment symmetries alone. It follows instead from the
quadrupolar truncation identity, cf.~Eq.~\eqref{eq:cons_vanish_2} $S_{i(j|kl)}=0$. Indeed,
$2S_{aj|al}=-S_{jl|aa}$, which is symmetric in $(j,l)$. The $l=1$ magnetic
projection of the tensor sector therefore vanishes only once the quadrupolar truncation
 is imposed.

Substituting~\eqref{eq:l1_chi0i} into~\eqref{huA}--\eqref{hAB} yields the
$l=1$ magnetic contributions
\begin{align}
h_{uA}\,\big|_{X^{l=1}_A} &= -\frac{2}{r}\,J_i\,X^{\langle i\rangle}_A ,
  \label{eq:l1_huA}\\[2pt]
h_{rA}\,\big|_{X^{l=1}_A} &= -\frac{2}{r\,(1+Hr)}\,J_i\,X^{\langle i\rangle}_A ,
  \label{eq:l1_hrA}\\[2pt]
h_{AB}\,\big|_{X^{l=1}_A} &= 0.
  \label{eq:l1_hAB}
\end{align}
Written out with $X^{\langle i\rangle}_A=\epsilon_{ijk}\,n^j e^k_A$, the
component~\eqref{eq:l1_huA} is independent of $H$ and reproduces the
magnetic-parity dipole of CHK~\cite{Compere:2023ktn}, equivalently Eq.~(5.38)
of~\cite{Harsh:2024kcl}, up to the gauge transformation relating the
generalised harmonic and Bondi gauges.\footnote{The radial
component~\eqref{eq:l1_hrA} is specific to the generalised harmonic gauge
adopted here and is removed by the transformation to Bondi gauge.}

Since both surviving components decay at least as $1/r$, they vanish at
$\mathcal{I}^+$ and therefore leave no radiative imprint.
The metric perturbation~\eqref{eq:l1_huA}--\eqref{eq:l1_hAB} satisfy the linearised Einstein
equations.

\subsection{Electric sector}

The electric sector differs from its magnetic counterpart in two respects.
First, the scalar harmonics now participate: all six metric components
contribute, and the extraction involves seven structures rather than three,
\begin{align}
& \hat{\chi}, & & n^i\chi_{0i}, & & n^in^j\chi_{ij}, & & \delta^{ij}\chi_{ij},
& & e^i_A\,\chi_{0i}, & & n^ie^j_A\,\chi_{ij}, & & e^i_{\langle A}e^j_{B\rangle}\,\chi_{ij}.
\label{eq:el2_structures}
\end{align}
The first four enter $h_{uu}$, $h_{ur}$, $h_{rr}$, and the trace part of
$h_{AB}$ through their scalar-harmonic coefficients, proportional to
$\widehat n^{ij}$, $n^i$, and $1$. The next two enter $h_{uA}$ and $h_{rA}$
through the electric vector harmonics $Y^{\langle ij\rangle}_A$ and
$Y^{\langle i\rangle}_A$. The last contributes to the trace-free part of
$h_{AB}$ through the electric tensor harmonic $Y^{\langle ij\rangle}_{AB}$.

Second, the extraction requires  trace identities among the quadrupolar moments, which we derive first. They follow from the truncation condition $S_{i(j|kl)}=0$,
Eq.~\eqref{eq:cons_vanish_2}, under two contractions. Contracting the first index with one index of the symmetrised group gives
\be
\Qp_{kl} + S_{ak|al} + S_{al|ak} = 0,
\ee
while contracting the last two indices gives $2S_{ia|aj} = -S_{ij|aa}$. Since
the right-hand side of the latter is symmetric in $(ij)$, so is $S_{ia|aj}$. The two
relations combine to give
\be
S_{ij|kk} = \Qp_{ij}, \qquad\qquad S_{ai|aj} = -\frac{1}{2}\,\Qp_{ij}.
\label{eq:el2_trace_identities}
\ee
Hence, under the quadrupolar truncation, every trace contraction of
$S_{ij|kl}$ reduces to the pressure quadrupole $\Qp_{ij}$. 

With these expressions at hand, another useful way of writing the tensor solution $\chi_{ij}$ is as follows. Let us define
\be
C_{ij} := 
- 4H^2 P_{(i|j)}
+ \frac{2}{3}H^2\partial_u\Qp_{ij} - 2H^3\Qp_{ij}.
\label{eq:el2_Cij_def}
\ee
With this, the tensor
solution~\eqref{eq:chi_ij_final} organises by STF angular content as
\bea
\chi_{ij} &=& \left(1 + \frac{1}{r}H^{-2}\partial_u\right) C_{ij}
+ \left[4H^2 + \frac{4}{r}\partial_u + \frac{4}{r^2}\right] n_k S_{ij|k} \nonumber \\[2pt]
&& + \left(2H^2\partial_u + \frac{2}{r}\,\partial_u^2 + \frac{6}{r^2}\,\partial_u
+ \frac{6}{r^3}\right)\widehat n_{kl} S_{ij|kl}.
\label{eq:el2_chi_organised}
\eea
Together, the three groupings carry $l\le 2$ scalar, $l\le 2$ vector, and
$l=2$ tensor content, and are projected mode by mode in the subsections that follow.

\subsubsection{\texorpdfstring{$l=2$}{l=2} terms}
 The scalar and vector
structures are read off from the equations~\eqref{eq:sec3-chihat-quad-operator-form}--\eqref{eq:sec3-chi0i-quad-operator-form},
and the $\chi_{ij}$ contractions from~\eqref{eq:el2_chi_organised}.

\paragraph{Scalar structures.} These require only the peeling
formula~\eqref{peeling-formula}. From
\eqref{eq:sec3-chihat-quad-operator-form}, collecting the coefficients of
$\widehat n^{ij}$, we have
\be
\hat{\chi}\,\Big|_{l=2}
= \left[\frac{2}{r}(\partial_u^2 - H^2) + \frac{6}{r^2}\,\partial_u + \frac{6}{r^3}\right]
\Qrp_{\langle ij\rangle}\,\widehat n^{ij}.
\label{eq:el2_chihat}
\ee
For $n^i\chi_{0i}$, the form
\eqref{eq:sec3-chi0i-quad-operator-form} shows that the $l=2$ scalar content
comes only from the one normal vector term.  We have,
\be
n^i\chi_{0i}\,\Big|_{l=2}
= 4\left[\frac{1}{r}\,\partial_u + \frac{1}{r^2}\right]
P_{\langle i|j\rangle}\,\widehat n^{ij}.
\label{eq:el2_nchi0i}
\ee
The trace of the tensor sector follows from
\eqref{eq:el2_chi_organised}. The
$l=2$ content comes entirely from the last term via $S_{ii|kl}=\Qp_{kl}$,
\be
\delta^{ij}\chi_{ij}\,\Big|_{l=2}
= \left(2H^2\partial_u + \frac{2}{r}\,\partial_u^2 + \frac{6}{r^2}\,\partial_u
+ \frac{6}{r^3}\right)\Qp_{\langle ij\rangle}\,\widehat n^{ij}.
\label{eq:el2_trchi}
\ee
Finally, in the $n^in^j\chi_{ij}$  structure, the $n^in^jn^kS_{ij|k}$ term drops by \eqref{eq:S-symmetric-zero}, and the
$n^i n^j \widehat n_{kl}S_{ij|kl}$ term contributes only through
its trace pieces. The result is
\be
n^in^j\chi_{ij}\,\Big|_{l=2}
= \left\{\left(1 + \frac{1}{r}H^{-2}\partial_u\right) C_{\langle ij\rangle}
- \frac{1}{3}\left(2H^2\partial_u + \frac{2}{r}\,\partial_u^2 + \frac{6}{r^2}\,\partial_u
+ \frac{6}{r^3}\right)\Qp_{\langle ij\rangle}\right\}
\widehat n^{ij}.
\label{eq:el2_nnchi}
\ee

\paragraph{Vector structures.} As in the magnetic sector, the projection
identity is \eqref{final_C19}. We now need to read off its electric sector terms. For an arbitrary rank-2 tensor $T_{ik}$, the coefficient of
$Y^{\langle ik\rangle}_A$ in $e^i_A n^k T_{ik}$ is
$\frac{1}{2}T_{\langle ik\rangle}$. The $l=2$ electric content of
$e^i_A\chi_{0i}$ comes solely from the one-vector term in
\eqref{eq:sec3-chi0i-quad-operator-form}. This gives
\be
e^i_A\,\chi_{0i}\,\Big|_{Y^{l=2}_A}
= 2\left[\frac{1}{r}\,\partial_u + \frac{1}{r^2}\right]
P_{\langle i|j\rangle}\,Y^{\langle ij\rangle}_A,
\label{eq:el2_eAchi0i}
\ee
precisely half of \eqref{eq:el2_nchi0i}. For $n^ie^j_A\chi_{ij}$ we get, 
\be
n^ie^j_A\,\chi_{ij}\,\Big|_{Y^{l=2}_A}
= \left\{\frac{1}{2}\left(1 + \frac{1}{r}H^{-2}\partial_u\right) C_{\langle ij\rangle}
- \frac{1}{6}\left(2H^2\partial_u + \frac{2}{r}\,\partial_u^2 + \frac{6}{r^2}\,\partial_u
+ \frac{6}{r^3}\right)\Qp_{\langle ij\rangle}\right\}
Y^{\langle ij\rangle}_A,
\label{eq:el2_neAchi}
\ee
again exactly half of \eqref{eq:el2_nnchi}.

\paragraph{Tensor structure.} The structure
$e^i_{\langle A}e^j_{B\rangle}\chi_{ij}$ is decomposed using
Appendix~\ref{app:D}. A careful calculation gives
\be
e^i_{\langle A}e^j_{B\rangle}\,\chi_{ij}\,\Big|_{Y^{l=2}_{AB}}
= \left\{\frac{1}{2}\left(1 + \frac{1}{r}H^{-2}\partial_u\right) C_{\langle ij\rangle}
+ \frac{1}{3}\left(2H^2\partial_u + \frac{2}{r}\,\partial_u^2 + \frac{6}{r^2}\,\partial_u
+ \frac{6}{r^3}\right)\Qp_{\langle ij\rangle}\right\}
Y^{\langle ij\rangle}_{AB}.
\label{eq:el2_eeAB}
\ee

Substituting \eqref{eq:el2_chihat}--\eqref{eq:el2_eeAB} into $h_{\mu \nu}$ yields the purely $l=2$ contributions in the electric sector. Direct substitution verifies that these terms satisfy the linearised Einstein equations upon using the conservation equation \eqref{eq:Qrho_dot_l=2}.

\subsubsection{\texorpdfstring{$l=1$}{l=1} terms}
 
There are no electric tensor harmonics at $l=1$. The sector is carried by the four scalar structures and the
two vector structures. It is governed by the three dipole variables $\Qrp_i$,
$P_i$, and $P_{i|kk}$, whose dynamics follow from the conservation equations~\eqref{eq:Qrho_dot}--\eqref{eq:P_dot}: $\partial_u P_i = -HP_i$ at $l=0$, and
$\partial_u\Qrho_i = H(\Qrho_i-\Qp_i)-P_i$ at $l=1$.  As before, the relevant trace
identities follow from the truncation
conditions. Contracting $S_{(ij|k)}=0$ over
the first two indices gives $S_{ia|a}=-\tfrac12\Qp_i$, and contracting
$P_{(i|jk)}=0$  gives
$P_{a|ak}=-\tfrac12 P_{k|aa}$. Tracing the $l=2$
flux law~\eqref{eq:P_dot_l=2} then gives
\be
\Qp_i = (\partial_u - H)P_{i|kk},
\label{eq:el1_Qp_dipole}
\ee
which we use throughout to eliminate $\Qp_i$.
 
\paragraph{Scalar structures.}
From~\eqref{eq:sec3-chihat-quad-operator-form}, the terms linear in $n^i$ give
\be
\hat{\chi}\,\Big|_{l=1}
= 4\left[\frac{1}{r}\,\partial_u + \frac{1}{r^2}\right]\Qrp_i\; n^i.
\label{eq:el1_chihat}
\ee
For $n^i\chi_{0i}$, we find 
\be
n^i\chi_{0i}\,\Big|_{l=1}
= \left\{\frac{4}{r}\,P_i
- \frac{2H}{r}(\partial_u - H)P_{i|kk}
- \frac{2}{r^2}\,\partial_u P_{i|kk}
- \frac{2}{r^3}\,P_{i|kk}\right\} n^i.
\label{eq:el1_nchi0i}
\ee
In the form~\eqref{eq:el2_chi_organised} the longitudinal projection
of the tensor sector vanishes,
\be
n^in^j\chi_{ij}\,\Big|_{l=1} = 0,
\label{eq:el1_nnchi}
\ee
since $C_{ij}$ carries no $l=1$ content, $n^in^jn^k S_{ij|k}$ is totally
symmetric, and $\widehat n_{kl}S_{ij|kl}$ contributes only at even $l$. The trace
receives its $l=1$ content entirely from the $n_kS_{ij|k}$ term via
$S_{ii|k}=\Qp_k$,
\be
\delta^{ij}\chi_{ij}\,\Big|_{l=1}
= \left(4H^2 + \frac{4}{r}\,\partial_u + \frac{4}{r^2}\right)
(\partial_u - H)P_{i|kk}\; n^i.
\label{eq:el1_trchi}
\ee
 
\paragraph{Vector structures.}
For $e^i_A\chi_{0i}$, the angle-independent term
of~\eqref{eq:sec3-chi0i-quad-operator-form} projects onto
$Y^{\langle i\rangle}_A=e^i_A$ with unit coefficient, and the rank-two term
contributes through the $Y^{\langle L-1\rangle}_A$ piece
of~\eqref{final_C19} at $l=2$. Using~\eqref{eq:el1_Qp_dipole}, we find
\be
e^i_A\,\chi_{0i}\,\Big|_{Y^{l=1}_A}
= \left\{\frac{4}{r}\,P_i
+ \frac{1}{r}(\partial_u - H)^2P_{i|kk}
+ \frac{1}{r^2}\,\partial_u P_{i|kk}
+ \frac{1}{r^3}P_{i|kk}\right\} Y^{\langle i\rangle}_A.
\label{eq:el1_eAchi0i}
\ee
For the companion structure $n^ie^j_A\chi_{ij}$, the $l=1$ electric content
arises solely from the $n_kS_{ij|k}$ term of~\eqref{eq:el2_chi_organised}.
Using $S_{ia|a}=-\tfrac12\Qp_i$ and~\eqref{eq:el1_Qp_dipole}, we find
\be
n^ie^j_A\,\chi_{ij}\,\Big|_{Y^{l=1}_A}
= -\left[H^2 + \frac{1}{r}\,\partial_u + \frac{1}{r^2}\right]
(\partial_u - H)P_{i|kk}\; Y^{\langle i\rangle}_A.
\label{eq:el1_neAchi}
\ee
  
Substituting these structures into $h_{\mu \nu}$ yields the purely $l=1$ contributions in the electric sector. Direct substitution verifies that these terms satisfy the linearised Einstein equations. We note that both
\eqref{eq:el1_trchi} and \eqref{eq:el1_neAchi} carry terms independent of
$1/r$, proportional to $H^2(\dt-H)P_{i|kk}$. These terms, which enter
$h_{uu}$ and $h_{uA}$ through \eqref{huA}, are removable by a gauge transformation \cite{Compere:2023ktn, Harsh:2024kcl}.

\subsubsection{\texorpdfstring{$l=0$}{l=0} terms}
 
The $l=0$ sector, present only in the electric parity, is the simplest. With no vector or tensor harmonics, it is carried entirely by the four scalar
structures, and the extraction reduces to isolating the angle-independent
parts.  The sector is governed by the
monopole and the trace $P_{k|k}$. Tracing $S_{ij}=-\partial_u P_{(i|j)}$ with
$S_{kk}=\Qp$ gives $\Qp=-\partial_u P_{k|k}$, which combined with the $l=0$ mass-flux
law $\partial_u\Qrho=-H\Qp$, yields the conservation law
\be
\partial_u E = 0,
\qquad\qquad
E := \Qrho - H P_{k|k}.
\label{eq:el0_energy}
\ee
As the notation suggests, $E$ is the conserved energy of the quadrupolar truncation,
reducing to $\Qrho$ in the flat limit (see also section 5.1 of \cite{Harsh:2024kcl}). 
 
\paragraph{Scalar structures.}
From~\eqref{eq:sec3-chihat-quad-operator-form}, the angle-independent terms
reside entirely at order $1/r$, and the monopole and quadrupole-trace
contributions combine as
\be
\hat{\chi}\,\Big|_{l=0}
= \frac{4}{r}\left[\Qrp
+ \frac{1}{6}\,(\partial_u - H)(\partial_u - 2H)\,\Qrp_{kk}\right].
\label{eq:el0_chihat}
\ee
For $n^i\chi_{0i}$, only the trace part of $n^in_jP_{i|j}$
in~\eqref{eq:sec3-chi0i-quad-operator-form} survives at $l=0$, giving
\be
n^i\chi_{0i}\,\Big|_{l=0}
= \frac{4}{3}\left[\frac{1}{r}\,\partial_u + \frac{1}{r^2}\right] P_{k|k}.
\label{eq:el0_nchi0i}
\ee
 For the longitudinal projection $n^i n^j \chi_{ij}$ of the tensor sector, writing
$\widehat n_{kl}=n_kn_l-\tfrac13\delta_{kl}$ together with the trace identity
$S_{ij|kk}=\Qp_{ij}$ yields
\be
n_i n_j\, \widehat n_{kl} S_{ij|kl}
= -\frac{1}{3}\, n_i n_j\, \Qp_{ij}.
\label{eq:el0_collapse}
\ee
Its  $l=0$ part,
together with the trace of $C_{ij}$, gives
\be
n^in^j\chi_{ij}\,\Big|_{l=0}
= \frac{1}{3}\left\{\left(1 + \frac{1}{r}H^{-2}\partial_u\right) C_{kk}
- \frac{1}{3}\left(2H^2\partial_u + \frac{2}{r}\,\partial_u^2 + \frac{6}{r^2}\,\partial_u
+ \frac{6}{r^3}\right)\Qp_{kk}\right\}.
\label{eq:el0_nnchi}
\ee
Finally, from~\eqref{eq:el2_chi_organised}, we have, 
\be
\delta^{ij}\chi_{ij}\,\Big|_{l=0}
= \left(1 + \frac{1}{r}H^{-2}\partial_u\right) C_{kk},
\qquad
C_{kk} = - 4H^2 P_{k|k}
+ \frac{2}{3}H^2\partial_u\Qp_{kk} - 2H^3\Qp_{kk}.
\label{eq:el0_trchi}
\ee
The linearised Einstein equations are satisfied.

\section{Octupolar radiation in de Sitter spacetime}
\label{sec:Octupolar}
In this section, we study octupolar gravitational radiation in de Sitter 
spacetime. From a theoretical perspective, this extension is necessary to 
construct a controlled and consistent multipolar expansion. Although the 
quadrupolar truncation provides the leading radiative contribution, higher multipoles 
encode additional degrees of freedom associated with the spatial structure 
and internal dynamics of the source. The inclusion of octupolar (and higher) terms is essential for a complete characterisation of the asymptotic 
radiation field. In what follows, we work within the octupolar truncation, 
defined by the assumption that all moments with $l > 3$ vanish:
\begin{equation}
\int d^3x\; a^{l+1}\, T_{\mu\nu}\, x^L = 0,
\qquad \forall \quad l > 3.
\label{eq:octu_trunc}
\end{equation}
The structure of this section is as follows. In 
Section~\ref{sec:expansion_octupolar}, we develop the systematic expansion 
of the relevant expressions up to octupolar order. 
Sections~\ref{sec:octupolar_scalar_vector} and~\ref{sec:octupolar_tensor} 
treat the scalar/vector and tensor inhomogeneous solutions respectively.

\subsection{Expansions to octupolar order}
\label{sec:expansion_octupolar}
The octupolar truncation requires extending the geometric and source 
expansions of Section~\ref{sec:scalar_vector} to one higher order in the source size $d$.\footnote{More precisely, the octupolar truncation requires extending the 
expansions of Section~\ref{sec:quadrupolar} to one higher polynomial degree in $x'$. The remainder terms vanish identically under \eqref{eq:octu_trunc}. Since the source size $d$ does not appear in any of the final expressions, it is no more than a  useful bookkeeping device.} 
The expansions of $|\vec{x}-\vec{x}'|$ and its reciprocal, already given 
to quadrupolar order in Eqs.~\eqref{eq:r_expand}--\eqref{eq:r_expand_inv}, acquire one further term:
\begin{equation} \label{eq:xxp_oct}
|\vec{x}-\vec{x}'| = \rho\!\left(
1
- \frac{p}{\rho}
+ \frac{q}{2\rho^2}
+ \frac{pq}{2\rho^3}
+ O\!\left(\frac{d^4}{\rho^4}\right)
\right).
\end{equation}
\begin{equation}\label{eq:1_over_xxp_oct}
\frac{1}{|\vec{x}-\vec{x}'|} = \frac{1}{\rho}\!\left(
1
+ \frac{p}{\rho}
+ \frac{3p^2 - \rho'^2}{2\rho^2}
+ \frac{p(5p^2 - 3 \rho'^2)}{2\rho^3}
+ O\!\left(\frac{d^4}{\rho^4}\right)
\right).
\end{equation}
The new terms at order $d^3/\rho^3$ are proportional to $pq$ and 
$p(5p^2 - 3 \rho'^2)$, respectively. The latter is, up to normalisation, the 
octupolar Legendre polynomial $P_3$.
Similarly, the expansion of $\eta/(\eta-|\vec{x}-\vec{x}'|)$ 
acquires three new terms at combined order $d^3$:
\begin{equation} \label{eq:eta_eta_xxp_oct}
\frac{\eta}{\eta - |\vec{x}-\vec{x}'|} = \frac{\eta}{\eta_{\rm ret}}\!\left(
1
- \frac{p}{\eta_{\rm ret}}
+ \frac{p^2}{\eta_{\rm ret}^2}
+ \frac{q}{2\rho\,\eta_{\rm ret}}
- \frac{p^3}{\eta_{\rm ret}^3}
- \frac{pq}{\rho\,\eta_{\rm ret}^2}
+ \frac{pq}{2\rho^2\,\eta_{\rm ret}}
+ \cdots
\right).
\end{equation}
where the omitted terms are schematically of orders 
\be
O\!\left(\frac{d^4}{\eta_{\rm ret}^4}, \; \frac{d}{\rho}\frac{d^3}{\eta_{\rm ret}^3},\;   \frac{d^2}{\rho^2}\frac{d^2}{\eta_{\rm ret}^2}, \;   \frac{d^3}{\rho^3}\frac{d}{\eta_{\rm ret}}\right).
\ee

\subsection{Scalar and vector inhomogeneous solutions}
\label{sec:octupolar_scalar_vector}

The only change with respect to Section~\ref{sec:scalar_vector} is 
that the integrand must now be expanded to one higher order in the source 
size $d$. With the expansions of Section~\ref{sec:expansion_octupolar} in hand, we write the retarded solution \eqref{eq:chi_hat_ret} as
\begin{equation}
\hat{\chi} = \frac{4\eta}{\rho\,\eta_{\rm ret}}
\int d^3x'\,(G_0 + G_1 + G_2 + G_3),
\end{equation}
where $G_0$, $G_1$ and $G_2$ are unchanged from Eqs.~\eqref{eq:G0}--\eqref{eq:G2}, and 
the new octupolar block $G_3$ is obtained by collecting all contributions 
of combined order $d^3$. The Taylor expansion of $\hat{T}$ around $\eta_{\rm ret}$ 
carried to third order takes the form, 
\begin{equation}
\hat{T}(\eta-|\vec{x}-\vec{x}'|,\vec{x}')
= C_0+C_1+C_2+C_3+O(d^4),
\end{equation}
where $C_0$, $C_1$ and $C_2$ are as in Eqs.~\eqref{eq:C0}--\eqref{eq:C2}, and the new 
term is
\begin{equation}
C_3 = \frac{1}{6}p^3\hat{T}^{(3)}
-\frac{pq}{2\rho}\hat{T}^{(2)}
-\frac{pq}{2\rho^2}\hat{T}^{(1)}.
\end{equation}
Substituting $C_3$ into the product of the expansions and 
collecting all contributions of combined order $d^3$ in the source size, 
one obtains
\begin{align}
\nonumber  G_3 =\;
&\left[
\frac{p (5p^2-3\rho'^2)}{2\rho^3}
- \frac{ p (5p^2-3\rho'^2)}{2\rho^2\eta_\ret}
+ \frac{p(2p^2-\rho'^2)}{\rho\,\eta_\ret^2}
- \frac{p^3}{\eta_\ret^3}
\right]\hat{T}^{(0)} \\[3pt]
\nonumber  +\;&\left[
\frac{p(5p^2-3\rho'^2)}{2\rho^2}
- \frac{p(2p^2- \rho'^2)}{\rho\,\eta_\ret}
+ \frac{p^3}{\eta_\ret^2}
\right]\hat{T}^{(1)}\\[3pt]
\nonumber  +\;&\left[
\frac{p(2p^2-\rho'^2)}{2\rho}
- \frac{p^3}{2\eta_\ret}
\right]\hat{T}^{(2)}\!\\[3pt]
+\;& \frac{1}{6}\, p^3 \, \hat{T}^{(3)}. \label{eq:G3}
\end{align}

We begin with the $\hat{T}^{(3)}$ term at the bottom of $G_3$. Using $ p= n_k x'_k$, together with Eq.~\eqref{eq:prefactor} and the identity \eqref{eq:scalar_identity}, we have for the last term, 
\begin{equation}
\hat{\chi}\big|_{G_3,\,\hat{T}^{(3)}}
= \frac{2}{3r}  \,n_k n_l n_m
(\partial_t-4H)(\partial_t-3H)(\partial_t-2H)\,
Q^{(\rho+p)}_{klm}.
\end{equation}
For the $\hat{T}^{(2)}$ term, using $1/\rho = a_\ret (H + 1/r)$ and $-1/\eta_\ret = H a_\ret$, we have,
\begin{equation}
\hat{\chi}\big|_{G_3,\,\hat{T}^{(2)}}
= \frac{2}{r}(\partial_t-4H)(\partial_t-3H)
\left[
\left( 3 H + \frac{2}{r}\right)n_k n_l n_m Q^{(\rho+p)}_{klm}
-\left(H + \frac{1}{r}\right)n_k Q^{(\rho+p)}_{kll}
\right].
\end{equation}

For the $\hat{T}^{(1)}$ term, $G_3$ contains three 
terms, which produce two independent angular structures
with coefficients
\begin{equation}
\left(\frac{5}{2\rho^2} - \frac{2}{\rho\eta_{\rm ret}}
+ \frac{1}{\eta_{\rm ret}^2}\right)n^k n^l n^m x'_k x'_l x'_m
+\left(-\frac{3}{2\rho^2}
+ \frac{1}{\rho\eta_{\rm ret}}\right)n^k x'_k x'^2.
\end{equation}
Applying the identity \eqref{eq:scalar_identity} with $l = 3$ and $n = 1$ gives
\begin{align}
& \int d^3x'\; x'_k x'_l x'_m\, \hat{T}^{(1)}
= a_\ret^{-3}(\partial_t - 4H)\, Q^{(\rho+p)}_{klm},\\ 
& \int d^3x'\; x'_k x'^2\, \hat{T}^{(1)}
= a_\ret^{-3}(\partial_t - 4H)\, Q^{(\rho+p)}_{kll}.
\end{align}
Using
\begin{equation}
\frac{1}{\rho^2} = a_\ret^2\!\left(H + \frac{1}{r}\right)^2,
\qquad
\frac{1}{\rho\,\eta_{\rm ret}} = -a_\ret^2 H\!\left(H + \frac{1}{r}\right),
\qquad
\frac{1}{\eta_{\rm ret}^2} = H^2 a_\ret^2,
\end{equation}
and evaluating the coefficients of each angular structure, the $\hat{T}^{(1)}$ contribution reads
\begin{align}
\hat{\chi}\big|_{G_3,\,\hat{T}^{(1)}}
=& \frac{2}{r}(\partial_t - 4H)  \times \nonumber \\
& \left[ 
\left( 11H^2 + \frac{14H}{r} +  \frac{5}{r^2} \right)
n^k n^l n^m Q^{(\rho+p)}_{klm}  
-  \left( 5H^2 +  \frac{8H}{r} + \frac{3}{r^2}  \right)
n^k Q^{(\rho+p)}_{kll}
\right].
\label{eq:G3_T1}
\end{align}
This generates contributions at orders $1/r$, $1/r^2$ and $1/r^3$.

For the $\hat{T}^{(0)}$ term, 
coefficients of the 
the two independent angular structures in $G_3$  are 
\begin{align}
&  \left( \frac{5}{2\rho^3} - \frac{5}{2\rho^2\eta_{\rm ret}}
+ \frac{2}{\rho\,\eta_{\rm ret}^2} - \frac{1}{\eta_{\rm ret}^3}\right) n_k n_l n_m x'_k x'_l x'_m , \\
&  \left(-\frac{3}{2\rho^3} + \frac{3}{2\rho^2\eta_{\rm ret}}
- \frac{1}{\rho\,\eta_{\rm ret}^2} \right) n_k x'_k x'^2.
\end{align}
Applying the identity \eqref{eq:scalar_identity} with $l = 3$ and $n = 0$,  gives,
\begin{equation}
\int d^3x'\; x'_k x'_l x'_m\, \hat{T}^{(0)}
= a_\ret^{-4} Q^{(\rho+p)}_{klm},
\qquad
\int d^3x'\; x'_k x'^2\, \hat{T}^{(0)}
= a_\ret^{-4} Q^{(\rho+p)}_{kll}.
\end{equation}
Following the steps as for the previous terms,  and collecting by powers of $1/r$, we get,
\begin{align}
\hat{\chi}\big|_{G_3,\,\hat{T}^{(0)}}
=& \frac{4}{r}
\left[
\left(
8H^3 + \frac{29H^2}{2r} + \frac{10H}{r^2} + \frac{5}{2r^3}
\right)
n^k n^l n^m Q^{(\rho+p)}_{klm} \right. \nonumber \\
& 
\left.  \qquad \qquad -\left(
4H^3 + \frac{17H^2}{2r} + \frac{6H}{r^2} + \frac{3}{2r^3}
\right)
n^k Q^{(\rho+p)}_{kll}
\right].
\label{eq:G3_T0}
\end{align}
This generates contributions at orders $1/r$, $1/r^2$, $1/r^3$ and
$1/r^4$. 

After expanding the differential operators and collecting by derivative
order, extensive cancellations occur and the
result simplifies to
\begin{align}
\hat{\chi}\big|_{G_3}
&= \frac{2}{r}
\left[
\frac{1}{3} \left(\partial_t^3 - 4H^2\partial_t\right)
Q^{(\rho+p)}_{klm}  n_k n_l n_m 
- H\left(\partial_t^2 - 2H\partial_t\right)
Q^{(\rho+p)}_{kll} n_k
\right]
\notag \\
&+ \frac{2}{r^2}
\left[
\left(2\partial_t^2 - 3 H^2\right)
Q^{(\rho+p)}_{klm} n_k n_l n_m
- \left(\partial_t^2 + H\partial_t - 3 H^2\right)
Q^{(\rho+p)}_{kll} n_k
\right]
\notag \\
&+ \frac{2}{r^3}
\left[
5\, \dot Q^{(\rho+p)}_{klm} n_k n_l n_m
- 3\, \dot Q^{(\rho+p)}_{kll} n_k
\right]
\notag \\
&+ \frac{2}{r^4}
\left[
5\,  Q^{(\rho+p)}_{klm}  n_k n_l n_m
- 3\, Q^{(\rho+p)}_{kll} n_k 
\right].
\label{eq:G3_simplified}
\end{align}
The $1/r^3$
and $1/r^4$ terms involve no $H$-dependent factors. 

The expression for $\chi_{0i}$ is exactly analogous and is obtained by the replacement 
\be
Q^{(\rho+p)}_{klm} \longrightarrow P_{i|klm}.
\ee

\subsection{Tensor inhomogeneous solution}
\label{sec:octupolar_tensor}

The tensor sector is the most involved part of the octupolar extension, and  qualitatively the richest. As recalled in Section~\ref{sec:tensor}, the full particular solution decomposes into three contributions: the direct light cone term $\chi^{(I)}_{ij}$, the global tail $\chi^{(II)}_{ij}$, and the near-zone tail correction $\chi^{(III)}_{ij}$. A key observation, already established in Section~\ref{sec:tensor}, is that $\chi^{(II)}_{ij}$ is exact and independent of the multipolar truncation order. It therefore carries over unchanged to the octupolar case and requires no further treatment here. The new contributions at octupolar order arise from $\chi^{(I)}_{ij}$ and $\chi^{(III)}_{ij}$, which are treated in Sections~\ref{sec:3.3.1} and~\ref{sec:3.3.2} respectively.

\subsubsection{The direct light cone term \texorpdfstring{$\chi^{(I)}_{ij}$}{chi(I) ij}}
\label{sec:3.3.1}
The extension to octupolar order follows the same organisational scheme as
in Section~\ref{sec:tensor}.  
Substituting the expansion of $1/|\vec{x}-\vec{x}'|$ to octupolar order
via~\eqref{eq:1_over_xxp_oct} and the Taylor expansion of $T_{ij}$ to third order via
\begin{equation}\label{eq:Taylor_T_ij}
T_{ij}(\eta - |\vec{x}-\vec{x}'|, x')
= C_{0 \: ij} + C_{1 \: ij} + C_{2 \: ij} + C_{3 \: ij} + O(d^4),
\end{equation}
where
\begin{align}
C_{3\:ij} &= \frac{1}{6}p^3 \, T^{(3)}_{ij}  - \frac{pq}{2\rho} \, T^{(2)}_{ij} -\frac{pq}{2\rho^2} \,T^{(1)}_{ij},
\label{eq:C3_ij}
\end{align}
we write the expansion for Eq.~\eqref{eq:chiI_def}  as,
\begin{equation}
\chi^{(I)}_{ij} = \frac{4}{\rho}\int d^3x'\,(G_{0\:ij} + G_{1\:ij} + G_{2\:ij} + G_{3\:ij}), \label{eq:3.13}
\end{equation}
where $G_{0\,ij}$, $G_{1\,ij}$ and $G_{2\,ij}$ are unchanged from Eqs.~\eqref{eq:G0ij_1}--\eqref{eq:G2ij_1},  and the new octupolar block is
\begin{align}
G_{3\:ij} &= \frac{p\,  (5p^2-3\rho'^2)}{2\rho^3}\,T^{(0)}_{ij}
+ \frac{p \,  (5p^2-3\rho'^2)}{2\rho^2}\,T^{(1)}_{ij}
+ \frac{p\, (2p^2-\rho'^2)}{2\rho}\,T^{(2)}_{ij}        + \frac{1}{6} \, p^3 \,T^{(3)}_{ij}. \label{eq:3.14}
\end{align}
Substituting $p = n_k x'_k$ and converting the spatial integrals to
source moments via the identity~\eqref{eq:identity} with $l = 3$, the angular
factors $p^3$, $p(2p^2 - \rho'^2)$ and $p(5p^2 - 3\rho'^2)$ in~\eqref{eq:3.14}
give rise respectively to $n_kn_ln_m$,
$(2n_kn_ln_m - n_k\delta_{lm})$ and
$(5n_kn_ln_m - 3n_k\delta_{lm})$, while the four terms
corresponding to $n = 0, 1, 2, 3$ in the identity each contribute
a distinct differential operator acting on $S_{ij|klm}$.
The octupolar contribution to the direct light cone term is thus
\begin{align}
\chi^{(I)}_{ij} \Big{|}_{l=3}
&= \frac{2\,n_k n_l n_m}{3\,a_\ret \rho}
(\partial_t-4H)(\partial_t-3H)(\partial_t-2H)\,S_{ij|klm}
\nonumber\\[3pt]
&\quad
+ \frac{2(2n_k n_l n_m - n_k\delta_{lm})}{a_\ret^2\rho^2}
(\partial_t-4H)(\partial_t-3H)\,S_{ij|klm}
\nonumber\\[3pt]
&\quad
+ \frac{2(5n_k n_l n_m - 3\,n_k\delta_{lm})}{a_\ret^3\rho^3}
(\partial_t-4H)\,S_{ij|klm}
\nonumber\\[3pt]
&\quad
+ \frac{2(5n_k n_l n_m - 3\,n_k\delta_{lm})}{a_\ret^4\rho^4}
S_{ij|klm}. \label{eq:3.15}
\end{align}

The notation $\chi^{(I)}_{ij}\big|_{l=3}$ indicates that~\eqref{eq:3.15}
captures only the new octupolar contribution. The complete direct
light cone term at octupolar truncation is obtained by adding~\eqref{eq:3.15}
to the quadrupolar contributions already given in~\eqref{eq:G0ij}--\eqref{eq:G2ij}.  The angular tensor $(5n_kn_ln_m - 3n_k\delta_{lm})$  appears in both the
$1/(a_\ret^3\rho^3)$ and $1/(a_\ret^4\rho^4)$ terms of~\eqref{eq:3.15},
in direct analogy with the pattern observed at quadrupolar order
in~\eqref{eq:G2ij}, where the same angular structure
$(3n_kn_l - \delta_{kl})$ appears in both
the $1/(a_\ret^2\rho^2)$ and $1/(a_\ret^3\rho^3)$ terms.
All moments in~\eqref{eq:3.15} are evaluated at $\eta_{\mathrm{ret}}$.

\subsubsection{The near-zone tail correction \texorpdfstring{$\chi^{(III)}_{ij}$}{chi(III) ij}}

\label{sec:3.3.2}

The extension of $\chi^{(III)}_{ij}$ to octupolar order also proceeds in direct 
analogy with the quadrupolar treatment of Section~\ref{sec:tensor}. It requires two modifications: (i) the octupolar extension of $\delta$, cf.~\eqref{eq:delta}, (ii) the inclusion of the  $\eta'^2$ term in the integrand expansion, cf.~\eqref{eq:integrand_expansion}. From the octupolar extension of $|\vec{x}-\vec{x}'|$ given in Eq.~\eqref{eq:xxp_oct},
the near-zone integration range becomes
\begin{equation}
\delta = p - \frac{q}{2\rho} - \frac{pq}{2\rho^2} + O(d^4).
\end{equation}
At octupolar truncation, the Taylor expansion of the integrand in 
Eq.~\eqref{eq:chiIII_def} must be carried to one higher order in $\eta'$:
\begin{align}
\frac{1}{1+\eta'/\eta_{\rm ret}}\, & \partial_{\eta'}T_{ij}(\eta_{\rm ret}+\eta',\vec{x}') \nonumber \\
&= T^{(1)}_{ij}
+ \eta'\!\left(T^{(2)}_{ij}-\frac{T^{(1)}_{ij}}{\eta_{\rm ret}}\right)
+ \eta'^2\!\left(\frac{T^{(3)}_{ij}}{2}
-\frac{T^{(2)}_{ij}}{\eta_{\rm ret}}
+\frac{T^{(1)}_{ij}}{\eta_{\rm ret}^2}\right)
+ O(\eta'^3).
\end{align}
Integrating term by term over $\eta'\in[0,\delta]$ yields
\begin{align}
\int_0^\delta\! d\eta'\, & \frac{1}{1+\eta'/\eta_{\rm ret}}\,  \partial_{\eta'}T_{ij} \nonumber \\
& = T^{(1)}_{ij}\,\delta
+ \frac{1}{2} \delta^2  \left(T^{(2)}_{ij}-\frac{T^{(1)}_{ij}}{\eta_{\rm ret}}\right)
+ \frac{1}{3} \delta^3  \left(\frac{T^{(3)}_{ij}}{2}
-\frac{T^{(2)}_{ij}}{\eta_{\rm ret}}
+\frac{T^{(1)}_{ij}}{\eta_{\rm ret}^2}\right) + \cdots. \label{eq:chiIII_def_new}
\end{align}
At quadrupolar order only the first two terms contributed. The third term 
and the new $O(d^3)$ piece of $\delta$ together generate the three distinct 
octupolar contributions.

The first octupolar contribution arises from the $O(d^3)$ correction $-pq/(2\rho^2)$ to 
$\delta$ acting on the linear-in-$\delta$ piece. Using~\eqref{eq:identity} 
with $l=3, n=1$ and $1/\eta_{\rm ret} = -Ha_\ret$ in \eqref{eq:chiIII_def}, we get,
\begin{equation}
-4Ha_\ret \int d^3x'\left(-\frac{pq}{2\rho^2}\right)T^{(1)}_{ij}
= \frac{2H}{a_\ret^2\rho^2}\,n_k(\delta_{lm}-n_ln_m)
(\partial_t-4H)\,S_{ij|klm}.
\end{equation}
Another contribution arises from the $O(d^3)$ cross-term 
$ -pq/\rho$ in the quadratic-in-$\delta$ piece.
Using~\eqref{eq:identity} with $n=2$ to handle $T^{(2)}_{ij}$ and 
$n=1$ to handle $H a_\ret T^{(1)}_{ij}$, we have,
\begin{align}
-4Ha_\ret & \int d^3x'\,\frac{1}{2}\!\left(-\frac{pq}{\rho}\right)
\!\left(T^{(2)}_{ij}+Ha_\ret T^{(1)}_{ij}\right)  \nonumber\\
&= \frac{2H}{a_\ret \rho}\,n_k(\delta_{lm}-n_ln_m)    \,(\partial_t-4H)\bigl[(\partial_t-3H)+H\bigr]S_{ij|klm}
\nonumber\\
&  = \frac{2H}{a_\ret\rho}\,n_k(\delta_{lm}-n_ln_m)
(\partial_t-4H)(\partial_t-2H)\,S_{ij|klm}.
\end{align}
Finally, the last contribution arises from $p^3$ in the 
cubic-in-$\delta$ piece. Using~\eqref{eq:identity} with $n=1,2,3$ for the 
three terms in the cubic integrand, we have, 
\begin{align}
-4H a_\ret & \int d^3x'\,\frac{1}{3}p^3
\left(\frac{1}{2}T^{(3)}_{ij}+Ha_\ret T^{(2)}_{ij}+H^2a_\ret^2 T^{(1)}_{ij}\right)
\nonumber\\
&= -\frac{4H}{3}\,n_kn_ln_m\,(\partial_t-4H)
\left[\tfrac{1}{2}(\partial_t-3H)(\partial_t-2H)
+H(\partial_t-3H)+H^2\right]S_{ij|klm}
\nonumber\\
&= -\frac{2H}{3}\,n_kn_ln_m
\,(\partial_t-4H)(\partial_t-2H)(\partial_t-H)\,S_{ij|klm}.
\end{align}
Adding all terms gives the new octupolar contribution. It reads, \begin{align}
\chi^{(III)}_{ij}\Big|_{l= 3}
~=~ & \frac{2H}{a_\ret^2\rho^2}\,n_k(\delta_{lm}-n_ln_m)
(\partial_t-4H)\,S_{ij|klm}
\nonumber\\
&
+\frac{2H}{a_\ret\rho}\,n_k(\delta_{lm}-n_ln_m)
(\partial_t-4H)(\partial_t-2H)\,S_{ij|klm}
\nonumber\\
&
-\frac{2H}{3}\,n_kn_ln_m
(\partial_t-4H)(\partial_t-2H)(\partial_t-H)\,S_{ij|klm},
\label{eq:chiIII_oct}
\end{align}
where all moments are evaluated at $\eta_{\rm ret}$. The structure of~\eqref{eq:chiIII_oct} closely parallels the direct 
light cone result~\eqref{eq:3.15}. The first two terms share the angular tensor 
$n_k(\delta_{lm}-n_ln_m)$, appearing at orders $1/(a_\ret^2\rho^2)$ and 
$1/(a_\ret\rho)$ respectively, while the last term carries the fully longitudinal 
structure $n_k n_l n_m$.

\subsubsection{Adding the two contributions}

We now add the two octupolar contributions $\chi^{(I)}_{ij}\Big|_{l=3}$
and $\chi^{(III)}_{ij}\Big|_{l=3}$ and organise the result by powers
of $1/r$. For compactness we introduce the shorthands
\begin{equation}
\mathcal{S}_{ij} := n_k n_l n_m S_{ij|klm},
\qquad
\mathcal{T}_{ij} := n_k S_{ij|kll},
\label{def_curlyS_curlyT}
\end{equation}
and use $1/(a_\ret \rho) = H + 1/r$ throughout. After a careful expansion of
the differential operators and collection of terms, extensive cancellations
occur at every order in $1/r$. The combined result reads
\begin{align}
\chi_{ij}\Big|_{l=3}
&= \chi^{(0)}_{ij}\Big|_{l=3} + \frac{1}{r} H^{-2} \partial_t \chi^{(0)}_{ij}\Big|_{l=3} 
+ \frac{2}{r^2}
(\partial_t - H)  \left[
2(\partial_t + H) \,\mathcal{S}_{ij}
- (\partial_t + 2H)\,\mathcal{T}_{ij}
\right]
\notag \\
&
+ \frac{2}{r^3}
\partial_t \left[
5\,\mathcal{S}_{ij}
- 3\,\mathcal{T}_{ij}
\right]
+ \frac{2}{r^4}
\left[
5\,\mathcal{S}_{ij} - 3\,\mathcal{T}_{ij}
\right],
\label{eq:chi_I_III_oct}
\end{align}
where the boundary datum $\chi^{(0)}_{ij}\Big|_{l=3}$ is defined as
the $r^0$ contribution,
\begin{equation}
\chi^{(0)}_{ij}\big|_{l=3}
= \frac{2}{3}H^2(\partial_t - H)
\left[(\partial_t + H)\,\mathcal{S}_{ij}- 3  H\,\mathcal{T}_{ij}\right].
\label{eq:chi0_oct_compact}
\end{equation}
The relationship between the boundary datum and the 
$1/r$ coefficient, first observed at quadrupolar order in Eq.~\eqref{eq:chi_ij_final}, is thus preserved at octupolar order.

\section{Spherical harmonics decomposition of the octupolar perturbation}
\label{sec:SH_oct}

We now apply the spherical-harmonic decomposition developed at quadrupolar
order in Section~\ref{sec:SH_quad} to the octupolar perturbation. The
transformation to Bondi coordinates remains unchanged~\eqref{huu}--\eqref{hAB} and the STF projection identities of
Appendices~\ref{app:B}--\ref{app:D} are independent of the multipolar
truncation. Consequently, the sequence of steps summarised in
Figure~\ref{fig:quad-sh-flow-chart} applies here as well. One first identifies
the scalar, vector, and tensor projections entering the Bondi components,
works at fixed parity and fixed $l$, and then uses Appendix~\ref{app:B} for the
STF--scalar-harmonic correspondence, Appendix~\ref{app:C} for vector
projections, and Appendix~\ref{app:D} for tensor projections. The resulting
coefficients are reduced with the octupolar conservation identities before
the metric is reconstructed.

What changes is the source content of these projections. At octupolar order,
the harmonic spectrum extends to $l=3$, and the moments $\Qrp_{ijk}$,
$P_{i|jkl}$, and $S_{ij|klm}$, defined in
Eqs.~\eqref{eq:Qrho_def}--\eqref{eq:Qrp_def}, are retained. Since the octupolar
truncation sets only moments with $l>3$ to zero, traces of these $l=3$ moments
can be non-zero; the same traces vanish under quadrupolar truncation, where all
moments with $l>2$ are set to zero. Consequently, the octupolar solution
contains both new $l=3$ modes and trace-induced corrections to the $l=2,1$
magnetic and electric sectors. Apart from these corrections, their extraction
follows the quadrupolar procedure of Section~\ref{sec:SH_quad}. We therefore
focus here on the genuinely octupolar $l=3$ modes, treating the magnetic sector
in Section~\ref{sec:oct_magnetic} and the electric sector in
Section~\ref{sec:oct_electric}.

\subsection{The octupolar magnetic sector}
\label{sec:oct_magnetic}

In the magnetic sector, all three structures
\begin{align}
& e^{i}_{A}\,\chi_{0i}, & & n^{i}e^{j}_{A}\,\chi_{ij}, &
  & e^{i}_{\langle A}e^{j}_{B\rangle}\,\chi_{ij}
\end{align}
contribute. As at quadrupolar order, the calculation relies on the projection
identities of Appendices~\ref{app:B}--\ref{app:D}. After the conservation
equations are imposed, the entire sector is determined by the multipole moment
$J_{ijk}$.

Consider first $e^{i}_{A}\,\chi_{0i}$. Its $l=3$ contribution arises solely
from the term proportional to $n^k n^l n^m P_{i|klm}$ in $\chi_{0i}$. This
term is obtained from the corresponding $\Qrp_L$ term in $\hat{\chi}$ by the
replacement $\Qrp_L\to P_{i|L}$; see Eq.~\eqref{eq:G3_simplified}. The $l=3$
specialisation of identity~\eqref{final_C19} gives
\begin{equation}
e^i_A\,\nhat^{\langle klm\rangle}P_{i|klm}\,\big|_{X^{l=3}_A}
=-\frac{1}{4}\,\STF_{ijk}\!\left[\epsilon_{iab}\,P_{a|bjk}\right]
X^{\langle ijk\rangle}_A
\equiv -\frac{1}{4}\,J_{ijk}\,X^{\langle ijk\rangle}_A .
\label{eq:l3_projection}
\end{equation}
where, in analogy with the quadrupolar quantity $J_{ij}$, we define the
rank-three multipole moment
\begin{equation}
J_{ijk}:=\STF_{ijk}\!\big[\epsilon_{iab}\,P_{a|bjk}\big].
\label{eq:current_octupole}
\end{equation}
Thus, $J_{ijk}$ is the Cartesian coefficient of the magnetic vector harmonic.
The scalar and vector retarded solutions have the same coefficients at each
order in $1/r$. We may therefore read the four coefficients of the
triple-normal contraction directly from Eq.~\eqref{eq:G3_simplified} and
combine them with the projection~\eqref{eq:l3_projection}, obtaining
\begin{equation}
e^i_A\,\chi_{0i}\,\big|_{X^{l=3}_A}
=-\frac{1}{2}\left[
\frac{1}{3r}\big(\partial_u^3-4H^2\partial_u\big)
+\frac{1}{r^2}\big(2\partial_u^2-3H^2\big)
+\frac{5}{r^3}\,\partial_u
+\frac{5}{r^4}\right]J_{ijk}\,X^{\langle ijk\rangle}_A .
\label{eq:l3_chi0i}
\end{equation}
This is the octupolar analogue of the quadrupolar result, with one additional
power of $1/r$. In the flat limit, the leading $1/r$ term is
$-\tfrac{1}{6r}\,\partial_u^3 J_{ijk}$, exhibiting the
$\partial_u^{\,l}$ scaling of a multipole of order $l$.

Both remaining projections originate from the term $n^k n^l S_{ij|kl}$ in the
quadrupolar-form solution~\eqref{eq:chi_ij_final}. Its $l=3$ harmonic vanishes
under quadrupolar truncation but is present when the $l=3$ moments are retained.
Substituting the boundary datum~\eqref{eq:chi0_def} into the tensor
solution~\eqref{eq:chi_ij_final} and retaining the
$n^k n^l S_{ij|kl}$ terms gives
\begin{equation}
\chi_{ij}\big|_{n^k n^l S_{ij|kl}}
=\left[2H^2\partial_u+\frac{2}{r}\partial_u^2
+\frac{6}{r^2}\partial_u+\frac{6}{r^3}\right]n^k n^l S_{ij|kl}.
\label{eq:l3_chiij_collected}
\end{equation}

For $n^i e^j_A\chi_{ij}$, contraction with $n^i$ followed by the conservation
equation $S_{i(j|kl)}=-\tfrac13(\partial_u-2H)P_{i|jkl}$ yields the identity
\begin{equation}
n^i n^k n^l S_{ij|kl}
=-\frac13(\partial_u-2H)\,n^k n^l n^m P_{j|klm}.
\label{eq:l3_bridge}
\end{equation}
This identity reduces the tensor contraction to the same triple-normal
contraction that appeared in $\chi_{0i}$. Applying~\eqref{eq:l3_projection}
then gives
\begin{equation}
n^i e^j_A\chi_{ij}\big|_{X^{l=3}_A}
=\frac{1}{6}(\partial_u-2H)
\left[H^2\partial_u+\frac{\partial_u^2}{r}
+\frac{3\partial_u}{r^2}+\frac{3}{r^3}\right]
J_{ijk}\,X^{\langle ijk\rangle}_A .
\label{eq:l3_nchij}
\end{equation}

For $e^i_{\langle A}e^j_{B\rangle}\chi_{ij}$, the $H^{(+1)}$ term in the
tensor-projection identity~\eqref{final_D30} maps
$\nhat^{\langle kl\rangle}S_{ij|kl}$ onto the $l=3$ magnetic tensor harmonic.
Differentiating the definition~\eqref{eq:current_octupole} and applying the
$l=3$ conservation equation~\eqref{eq:P_dot} gives
\begin{equation}
(\partial_u-2H)\,J_{ijk}
=-2\,\STF_{ijk}[\epsilon_{iab}\,S_{aj|bk}].
\label{eq:octupole_conservation}
\end{equation}
The conservation equation~\eqref{eq:octupole_conservation} then yields
\begin{equation}
e^i_{\langle A}e^j_{B\rangle}\chi_{ij}\big|_{X^{l=3}_{AB}}
=\frac{1}{6}(\partial_u-2H)
\left[H^2\partial_u+\frac{\partial_u^2}{r}
+\frac{3\partial_u}{r^2}+\frac{3}{r^3}\right]
J_{ijk}\,X^{\langle ijk\rangle}_{AB},
\label{eq:l3_eechij}
\end{equation}
which carries the same differential operator as~\eqref{eq:l3_nchij}.

Substitution of~\eqref{eq:l3_chi0i}, \eqref{eq:l3_nchij},
and~\eqref{eq:l3_eechij} into $h_{uA}$, $h_{rA}$, and $h_{AB}$ gives the complete
$l=3$ magnetic perturbation. The resulting metric satisfies the linearised
Einstein equations, providing a non-trivial consistency check of the projection
and moments  obtained via the use of the truncation conditions and conservation equations.

\subsection{The octupolar electric sector}
\label{sec:oct_electric}

We next consider the electric sector at $l=3$. Because scalar harmonics now
participate, all seven projections listed in Eq.~\eqref{eq:el2_structures} contribute to
the six independent groups of metric components, exactly as in the
quadrupolar analysis of Section~\ref{sec:SH_quad}. Before performing the
projections, we use the conservation equations to reduce the symmetrised
moments entering the solution to the STF parts of the two rank-three
quantities $\Qrho_{ijk}$ and $\Qp_{ijk}$.

In analogy with the quadrupolar moment identity~\eqref{eq:Sij_identity}, it is useful to combine the $l=3$
conservation equations~\eqref{eq:Qrho_dot}--\eqref{eq:P_dot} to obtain
\be
P_{(i|jk)}
= -\frac{1}{3}\left(\partial_u\Qrho_{ijk}
- 3H\,\Qrho_{ijk} + H\,\Qp_{ijk}\right),
\label{eq:oct_el_Psym}
\ee
while the second conservation equation gives
\be
S_{(ij|k)} = -\frac{1}{2}\,(\partial_u- H)\,P_{(i|jk)},
\label{eq:oct_el_flux_package}
\ee
which are the rank-three counterparts of the quadrupolar moment relations used
in Section~\ref{sec:moments}. Unlike at quadrupolar truncation,
$S_{(ij|k)}$ does not vanish. Relative to the quadrupolar vector
solution~\eqref{eq:chi_0i_final}, retaining this term and using the rank-three
conservation equation~\eqref{eq:oct_el_flux_package} gives the correction
\be
\delta\chi_{0i}
= 3\left[\frac{1}{r}\big(\partial_u^2 - H^2\big)
+ \frac{3}{r^2}\,(\partial_u - H)\right] n^k n^l\, P_{(i|kl)},
\label{eq:delta_chi0i}
\ee
which contributes to the $l=3$ electric harmonics.

At octupolar truncation, the $l=4$ moment $P_{i|jklm}$ vanishes, conservation equation~\eqref{eq:S_asym} then gives
\be
S_{j(i|klm)} = 0
\label{eq:oct_el_rank5}
\ee
It replaces the quadrupolar truncation
identity~\eqref{eq:cons_vanish_2} in the present calculation and has two consequences
that will be used repeatedly. First, contracting the tensor
$\mathcal{S}_{ij}$ defined in Eq.~\eqref{def_curlyS_curlyT} with one unit normal gives
\be
n^i\mathcal{S}_{ij}=0.
\ee
Second, the fully symmetrised conservation identity~\eqref{eq:S_sym} gives
$S_{(ij|klm)}=0$ at octupolar truncation. Tracing this equation and using the
rank-five identity~\eqref{eq:oct_el_rank5} to eliminate the mixed traces yields
$3S_{(ij|k)ll}=\Qp_{ijk}$. Since $n^in^jn^k$ is fully symmetric, it follows that
\be
n^i n^j n^k\, S_{ij|kll}
= n^i n^j n^k\, S_{(ij|k)ll}
= \frac{1}{3}\, n^i n^j n^k\,\Qp_{ijk}.
\label{eq:oct_el_trace_collapse}
\ee
The part of $S_{ij|kll}$ that is not fully symmetric is not fixed by these
relations, but it cancels from every projection considered below. No
additional moment is therefore required.

For later use, denote by $\mathcal{P}$ the operator multiplying
$n_kS_{ij|k}$ in the organised quadrupolar tensor solution~\eqref{eq:el2_chi_organised}:
\be
\mathcal{P} = 4H^2 + \frac{4}{r}\partial_u + \frac{4}{r^2}.
\ee
The operators multiplying $\mathcal{T}_{ij}$ and $\mathcal{S}_{ij}$ in the
octupolar tensor solution~\eqref{eq:chi_I_III_oct} are, respectively,
\begin{align}
\mathcal{Q} &:= -2\left[H^3(\partial_u - H) + \frac{H}{r}\,\partial_u(\partial_u - H)
+ \frac{1}{r^2}(\partial_u - H)(\partial_u + 2H) + \frac{3}{r^3}\,\partial_u
+ \frac{3}{r^4}\right],
\label{eq:oct_el3_Q_op}\\[4pt]
\mathcal{R} &:= \frac{2}{3}H^2(\partial_u^2 - H^2)
+ \frac{2}{3r}\,\partial_u(\partial_u^2 - H^2)
+ \frac{4}{r^2}(\partial_u^2 - H^2) + \frac{10}{r^3}\,\partial_u + \frac{10}{r^4}.
\label{eq:oct_el3_R_op}
\end{align}

\paragraph{Scalar structures.} Extracting the STF $l=3$ part of the terms with
three unit-normal contractions in the octupolar scalar solution~\eqref{eq:G3_simplified} gives
\be
\hat{\chi}\,\Big|_{l=3}
= \left[\frac{2}{3r}\big(\partial_u^3 - 4H^2\partial_u\big)
+ \frac{2}{r^2}\big(2\partial_u^2 - 3H^2\big)
+ \frac{10}{r^3}\,\partial_u + \frac{10}{r^4}\right]
\Qrp_{\langle ijk\rangle}\,\widehat{n}^{\langle ijk\rangle}.
\label{eq:oct_el3_chihat}
\ee
For $n^i\chi_{0i}$, the vector and scalar solutions have the same coefficients
at each order in $1/r$. Combining the octupolar vector contribution with
$\delta\chi_{0i}$ expresses the result entirely in terms of
$P_{(i|jk)}$:
\be
n^i\chi_{0i}\,\Big|_{l=3}
= 2\left[\frac{1}{r}\big(\partial_u^2 - H^2\big)
+ \frac{3}{r^2}\,\partial_u + \frac{3}{r^3}\right]
P_{(i|jk)}\,\widehat{n}^{\langle ijk\rangle}.
\label{eq:oct_el3_nchi0i}
\ee
For $n^in^j\chi_{ij}$, the $\mathcal{S}_{ij}$ contribution vanishes because
$n^i\mathcal{S}_{ij}=0$. The $\mathcal{T}_{ij}$ contribution reduces
by the trace identity~\eqref{eq:oct_el_trace_collapse}, while the
$n_kS_{ij|k}$ contribution is reduced using the rank-three conservation
equation~\eqref{eq:oct_el_flux_package}. Together they give
\be
n^in^j\chi_{ij}\,\Big|_{l=3}
= \left\{-\frac{1}{2}\,\mathcal{P}\,(\partial_u - H)\,P_{(i|jk)}
+ \frac{1}{3}\,\mathcal{Q}\,\Qp_{\langle ijk\rangle}\right\}
\widehat{n}^{\langle ijk\rangle}.
\label{eq:oct_el3_nnchi}
\ee
The $l=3$ part of the trace follows directly from $S_{aa|L}=\Qp_L$:
\be
\delta^{ij}\chi_{ij}\,\Big|_{l=3}
= \mathcal{R}\,\Qp_{\langle ijk\rangle}\,\widehat{n}^{\langle ijk\rangle}.
\label{eq:oct_el3_trchi}
\ee

\paragraph{Vector structures.} The vector-projection identity~\eqref{final_C19} shows that each
vector projection is exactly one third of the corresponding scalar contraction
in the scalar projection formulas~\eqref{eq:oct_el3_nchi0i}
and~\eqref{eq:oct_el3_nnchi}.

\paragraph{Tensor structure.} The $l=3$ part of
$e^i_{\langle A}e^j_{B\rangle}\chi_{ij}$ is obtained using the
tensor-projection formulas of Appendix~\ref{app:D}. It receives two
contributions. The first comes from the $H^{(+2)}$ term in the contraction
formula and contains the $n_kS_{ij|k}$ term and the trace-induced
$\mathcal{T}_{ij}$ term. Decomposing the product of three unit normals into
its STF and trace parts produces an additional $\mathcal{T}_{ij}$
contribution, which is needed to obtain the final coefficient of
$\mathcal{R}$.

The second contribution comes from the $H^{(0)}$ term in the
tensor-projection identity~\eqref{final_D31} and contains the STF part of
$\mathcal{S}_{ij}$. Using the rank-five identity~\eqref{eq:oct_el_rank5} and
the trace identity~\eqref{eq:oct_el_trace_collapse}, its coefficient reduces
to
\[
\STF_{jkl}[\delta_{im}\,S_{\langle ij\rangle\langle klm\rangle}]
=-\frac{4}{5}\,\Qp_{\langle jkl\rangle}.
\]
Combining these contributions gives
\be
e^i_{\langle A}e^j_{B\rangle}\,\chi_{ij}\,\Big|_{Y^{l=3}_{AB}}
= \left\{-\frac{1}{12}\,\mathcal{P}\,(\partial_u - H)\,P_{(i|jk)}
+ \frac{1}{18}\,\mathcal{Q}\,\Qp_{\langle ijk\rangle}
+ \frac{1}{6}\,\mathcal{R}\,\Qp_{\langle ijk\rangle}\right\}
Y^{\langle ijk\rangle}_{AB}.
\label{eq:oct_el3_eeAB}
\ee

Substituting the projected solutions
\eqref{eq:oct_el3_chihat}--\eqref{eq:oct_el3_eeAB} into the
reconstruction formulas~\eqref{huu}--\eqref{hAB} gives the complete $l=3$
electric perturbation. After the conservation equations
\eqref{eq:Qrho_dot}--\eqref{eq:P_dot} are imposed, the resulting metric
satisfies the linearised Einstein equations.

\section{Conclusions}
\label{sec:conclusions}

In this paper, we derived the gravitational field generated by a localised
matter source in de~Sitter spacetime at octupolar order in the multipolar
expansion, working in generalised harmonic gauge in the future Poincar\'e
patch. The scalar, vector, and tensor perturbation equations were solved
explicitly, including the cosmological tail, providing the first extension of
the de~Sitter multipolar expansion beyond quadrupolar order. At both the quadrupolar and octupolar orders, the truncation is the only approximation made. In a given truncation (specified by the polynomial degree in the source size $d$), outside the source the solutions are exact in $H$, in the source velocity and in $1/r$. In this
 respect the de Sitter construction is similar to the exact
   Damour-Iyer exterior solutions \cite{Damour:1990gj}.

In Section~\ref{sec:SH_quad}, we translated the quadrupolar
generalised harmonic gauge solution for $\hat\chi$, $\chi_{0i}$, and $\chi_{ij}$
from STF tensor form into a spherical-harmonic form adapted to the
Bondi-type coordinates $(u,r,\theta,\phi)$. Organising the perturbation by
parity and multipole order, we extracted the spherical-harmonic coefficients
of the metric components $h_{\mu\nu}$ mode by mode, using the projection identities of
Appendices~\ref{app:B}--\ref{app:D}. The magnetic sector is governed at $l=1$
by the conserved angular momentum $J_i$, and at $l=0$ the electric sector
yields the conserved combination $E = Q^{(\rho)} - H P_{k|k}$, the
de~Sitter version of the source energy. As a check on the construction, we
verified that the reassembled perturbation satisfies the linearised Einstein
equations independently at each parity and each $l$.

The octupolar solution of Section~\ref{sec:Octupolar} admits the same
decomposition, and the procedure of Section~\ref{sec:SH_quad} carries over
without modification. The projection identities of
Appendices~\ref{app:B}--\ref{app:D} hold at every multipole order, and the
octupolar truncation supplies the analogous trace relations among the source
moments. In Section~\ref{sec:SH_oct} we therefore present only the genuinely
octupolar $l=3$ modes, in both the electric and the magnetic sector.

The lower-$l$ content of the perturbation in the octupolar truncation is extracted by precisely
the steps used at quadrupolar order, but the outcome is not simply the
quadrupolar solution. Imposing the octupolar truncation does not merely add
$l=3$ content on top of an unchanged quadrupolar field. It also dresses the
lower-multipole sectors. We have carried out this extraction in full. The
entire effect on the lower sectors is governed by a single rank-two object, the
trace $P_{i|jkk}$. Indeed, the $l=4$ specialisation
of the flux law~\eqref{eq:Qrho_dot} gives $P_{(i|jkl)}=0$ at octupolar
truncation, which implies $P_{a|aij}=-P_{(i|j)kk}$ and $P_{a|akk}=0$, i.e., the trace part of
$P_{i|jkk}$ vanishes identically. Its antisymmetric part dresses the $l=1$
magnetic terms, and its symmetric trace-free part the $l=2$ electric terms. The
mechanism is the failure, at octupolar truncation, of the trace identities among
the moments on which the quadrupolar extraction relied. Correspondingly, the
effect is $O(H^2)$ and vanishes in the flat-space limit, as it must, since at
linear order an octupole moment cannot correct quadrupolar or dipolar terms in
Minkowski spacetime. As this dressing is a technical rather than a conceptual
novelty, and as its full presentation would lengthen an already long paper, we do
not reproduce the details here. We will present these details elsewhere. Whether the dressing terms that survive at $\mathcal{I}^+$ constitute
genuine radiative content for $\Lambda>0$, or are removable by the residual gauge
freedom, will also be settled there.

\bigskip

\paragraph{Acknowledgements:} 
We thank Abhay Ashtekar, Geoffrey Comp\`ere, Ghanashyam Date, Jahanur Hoque, and Kostas Skenderis  for discussions. We are grateful to Thomas B\"ackdahl for suggesting the use of \texttt{SymManipulator}\footnote{https://xact.es/SymManipulator/}, his package for \texttt{xAct}. We also thank Jahanur Hoque, Geoffrey Comp\`ere, and Ghanashyam Date for comments on an earlier version of the draft.  The work of A.V.\ was partly supported by SERB Core Research Grant CRG/2023/000545. O.S. is supported by SONATA BIS
grant 2023/50/E/ST2/00231 from the Polish National Science Centre and acknowledges his debt to the people of Poland for their steady and generous support of research in basic sciences.
We gratefully acknowledge the use of AI-assisted tools, particularly Claude and ChatGPT for help with fixing various coefficients in appendices \ref{app:C} and \ref{app:D}. We also acknowledge the support from the Infosys Foundation to CMI.

\appendix 

\section{A universal multipole moment differentiation identity}
\label{proof_identity}
We prove the identity
\begin{equation}\label{eq:moment_identity}
\int d^3x'\,x'^L\,T^{(n)}_{ij} 
= a^{n-l-1}
\prod_{k=0}^{n-1}\!\bigl(\partial_t - (l+1-k)H\bigr)\,S_{ij|L} 
\end{equation}
by induction on $n$, where 
$T^{(n)}_{ij} := \partial^n_\eta T_{ij}$. Let us start with the base case $n=0$.
By definition of the stress-energy moments,
\begin{equation}
S_{ij|L} = \int d^3x'\,a^{l+1}\,x'^L \,T_{ij},
\end{equation}
so that
\begin{equation}
\int d^3x'\,x'^L\,T^{(0)}_{ij} = a^{-l-1}S_{ij|L},
\end{equation}
which matches the right-hand side of \eqref{eq:moment_identity} for $n=0$
(empty product equals 1). We now turn to the inductive step.
Assume the identity holds for $n-1$:
\begin{equation}\label{eq:ind_hyp}
\int d^3x'\,x'^L\,T^{(n-1)}_{ij}
= a^{n-l-2}
\prod_{k=0}^{n-2}\!\bigl(\partial_t-(l+1-k)H\bigr)\,S_{ij|L}.
\end{equation}
Differentiating both sides with respect to $\eta$ and using
$\partial_\eta = a\,\partial_t$ gives
\begin{equation}
\int d^3x'\,x'^L\,T^{(n)}_{ij}
= a\,\partial_t\!\left[
a^{n-l-2}
\prod_{k=0}^{n-2}\!\bigl(\partial_t-(l+1-k)H\bigr)S_{ij|L}
\right].
\end{equation}
Applying the Leibniz rule $a\,\partial_t(a^m f) = a^{m+1}[(\partial_t + mH)f]$
with $m = n-l-2$:
\begin{equation}
= a^{n-l-1}\bigl(\partial_t + (n-l-2)H\bigr)
\prod_{k=0}^{n-2}\!\bigl(\partial_t-(l+1-k)H\bigr)S_{ij|L}.
\end{equation}
Recognising that $\partial_t + (n-l-2)H = \partial_t - (l+2-n)H$,
and noting that all operators $(\partial_t - cH)$ with constant $c$ commute
with one another since
\begin{equation}
\bigl[(\partial_t - aH),(\partial_t - bH)\bigr]f = 0
\quad \forall\, a,b \in \mathbb{R},
\end{equation}
the new factor $(\partial_t - (l+2-n)H)$ can be absorbed into the
product to give
\begin{equation}
\int d^3x'\,x'^L\,T^{(n)}_{ij}
= a^{n-l-1}
\prod_{k=0}^{n-1}\!\bigl(\partial_t - (l+1-k)H\bigr)\,S_{ij|L},
\end{equation}
which is precisely \eqref{eq:moment_identity} for $n$.

Since the proof relies only on the definition of the moments and the 
relation $\partial_\eta = a\,\partial_t$, it carries over verbatim to 
the scalar and vector sectors, with the moment $S_{ij|L}$ replaced by 
the appropriate sector moment in each case.  For the scalar sector,  recall the definition of the scalar moment
\begin{equation}
Q^{(\rho+p)}_L := \int d^3x'\,a^{l+1}\,x'^L\,\hat{T},
\qquad \hat{T} := T_{00} + \delta^{ij}T_{ij}.
\end{equation}
The identity reads
\begin{equation}\label{eq:scalar_identity}
\int d^3x'\,x'^L\,\hat{T}^{(n)}
= a^{n-l-1}
\prod_{k=0}^{n-1}\!\bigl(\partial_t - (l+1-k)H\bigr)\,
Q^{(\rho+p)}_L,
\end{equation}
where $\hat{T}^{(n)} := \partial^n_\eta \hat{T}$.

For the vector sector,  recall the definition of the vector moment
\begin{equation}
P_{i|L} := \int d^3x'\,a^{l+1}\,x'^L\,T_{0i}.
\end{equation}
The identity reads
\begin{equation}\label{eq:vector_identity}
\int d^3x'\,x'^L\,T^{(n)}_{0i} 
= a^{n-l-1}
\prod_{k=0}^{n-1}\!\bigl(\partial_t - (l+1-k)H\bigr)\,
P_{i|L} ,
\end{equation}
where $T^{(n)}_{0i} := \partial^n_\eta T_{0i}$.

All three identities --- scalar \eqref{eq:scalar_identity}, vector
\eqref{eq:vector_identity}, and tensor \eqref{eq:moment_identity} ---
share an identical structure. The only difference is the replacement
\begin{equation}
Q^{(\rho+p)}_L \;\longleftrightarrow\; P_{i|L}
\;\longleftrightarrow\; S_{ij|L}
\end{equation}
of the relevant moment. In particular, the differential operator
$a^{n-l-1}\prod_{k=0}^{n-1}(\partial_t-(l+1-k)H)$ is
universal across all three sectors, depending only on the multipolar
order $l$ and the number of time derivatives $n$, and reducing
to $\partial^n_t$ in the flat limit $H\to 0$.

\section{STF formalism and its translation to spherical harmonics}
\label{app:B}
This appendix summarizes the elements of the symmetric trace-free (STF) formalism and its relation to vector and tensor spherical harmonics that are required in this work. The STF language provides a natural and efficient framework for describing higher multipole moments, and is particularly well-suited for deriving the octupolar extensions of the results in \cite{Compere:2023ktn}. For comparison with the broader literature \cite{Bonga:2023eml, Harsh:2024kcl}, as well as for explicitly verifying the Einstein equations, it is necessary  to translate the STF expressions into  vector and tensor spherical harmonics.

\subsection*{Spherical harmonics in STF form}

To set the notation, we begin by summarising  the relation between STF tensors and spherical harmonics, following \cite{poisson2014gravity}. The two formalisms provide equivalent decompositions of angular dependence, but are adapted to different coordinate choices. The decomposition in spherical harmonics relies on spherical polar coordinates and separates the angular variables $(\theta,\phi)$ from the radial coordinate. In contrast, the STF decomposition is formulated directly in Cartesian coordinates, treating all spatial directions on an equal footing.

The basic building blocks of the STF formalism are products of the unit radial vector $n^i$, projected onto their symmetric trace-free part.\footnote{Three-dimensional Cartesian indices are raised and lowered using $\delta_{ij}$; their position carries no significance.} Explicitly, the first two STF combinations are
\begin{align}
\widehat n^{\langle ij \rangle} &= n^i n^j - \frac{1}{3}\delta^{ij}, \\
\widehat n^{\langle ijk \rangle} &= n^i n^j n^k - \frac{1}{5}\left(\delta^{ij} n^k + \delta^{ik} n^j + \delta^{jk} n^i\right), 
\end{align}
where angular brackets (and the hat) denote STF projection. These expressions are constructed by symmetrizing tensor products of $n^i$ and subtracting all traces. The general STF projection of a product of $l$ unit vectors is given by~\cite{Blanchet:1985sp}
\begin{equation}
\widehat n^{\langle j_1 j_2 \cdots j_l \rangle}
=
\sum_{p=0}^{\lfloor l/2 \rfloor}
\frac{(-1)^p l!(2l - 2p - 1)!!}{(l - 2p)!(2l - 1)!!(2p)!!}
\,\delta^{(j_1 j_2} \cdots \delta^{j_{2p-1} j_{2p}} 
n^{j_{2p+1}} \cdots n^{j_l)}, 
\label{1.154}
\end{equation}
where parentheses denote symmetrization over all indices. This expression systematically removes all traces and produces an irreducible rank-$l$ Cartesian tensor.

The key result is that STF tensors provide an alternative representation of the same irreducible angular structures that appear in spherical harmonics. This correspondence becomes explicit once the unit vector is written in spherical coordinates,
\begin{equation}
n_i = (\sin\theta\cos\phi,\, \sin\theta\sin\phi,\, \cos\theta). \label{1.163}
\end{equation}
The STF tensor $\widehat n^{\langle L \rangle}$ (with multi-index $L$ of length $l$) is then a function on the sphere that transforms irreducibly under rotations, and therefore admits an expansion in spherical harmonics with fixed $l$:
\begin{equation}
\widehat n^{\langle L \rangle}
=
N_l \sum_{m=-l}^{l}
\mathcal{Y}^{\langle L \rangle}_{l m} \, Y_{l m}(\theta,\phi),
\qquad
N_l = \frac{4\pi l!}{(2l+1)!!}.
\label{1.164}
\end{equation}
Here $\mathcal{Y}^{\langle L \rangle}_{l m}$ are constant STF tensors that encode the mapping between the Cartesian STF basis and the spherical harmonic basis. The inverse relation expresses spherical harmonics directly in terms of STF tensors:
\begin{equation}
Y_{l m}(\theta,\phi)
=
\mathcal{Y}^{*\langle L \rangle}_{l m} \, \widehat n_{\langle L \rangle}.
\label{1.167}
\end{equation}
Equations (\ref{1.164}) and (\ref{1.167}) establish a one-to-one correspondence between STF tensors and  spherical harmonics. The normalisation $N_l$ was chosen so as to make the numerical factor on the right hand side of \eqref{1.167} equal to unity. Box 1.5 of \cite{poisson2014gravity} displays $\mathcal{Y}^{\langle L \rangle}_{l m} $ for selected smaller values of $l$.

\subsection*{Vector harmonics}

Vector spherical harmonics can be constructed from scalar harmonics by applying
covariant derivatives on the unit sphere and by forming axial combinations. Let $D_A$ denote the covariant derivative on the unit sphere, and let $ e_{Ai} = D_A n_i. $ The electric-type vector harmonic is defined by
\begin{equation}
Y_A^{l m} := D_A Y_{l m}.
\end{equation}
Using the STF representation, we obtain
\begin{equation}
Y_A^{l m}
=
\mathcal{Y}^{*\langle L \rangle}_{l m} \, D_A \widehat n_{\langle L \rangle}
= l \, \mathcal{Y}^{*\langle L \rangle}_{l m}  e_{A \langle i_l} \widehat n_{L-1 \rangle}.
\end{equation}

The magnetic-type vector harmonic is defined using the Levi--Civita tensor
$\epsilon_{AB}$ on the unit sphere:
\begin{equation}
X_A^{l m} := \epsilon_A{}^{B} D_B Y_{l m}.
\end{equation}
The Levi--Civita tensor on the unit two-sphere is defined so that $\epsilon_{\theta\phi} = \sin\theta$. In STF form this becomes
\begin{equation}
X_A^{l m}
=
l \, \mathcal{Y}^{*\langle L \rangle}_{l m} \,
\epsilon_A{}^B e_{B \langle i_l} \widehat n_{L-1 \rangle}.
\label{A.12}
\end{equation}
It is useful to express the right hand side of the above equation in terms of the three-dimensional Levi--Civita tensor. 
Using the identity 
\begin{equation}
\epsilon_{AB} = \epsilon_{ijk}\, e_{A i} e_{Bj} n_k,
\end{equation}
we have
\begin{equation}
\epsilon_A{}^{B} e_{B i} = \epsilon_{ijk}\, n_j e_{Ak}.
\end{equation}
Substituting this into Eq.~(\ref{A.12}), we  can write 
\begin{equation}
X_A^{l m}
= l \, \mathcal{Y}^{*\langle L \rangle}_{l m} \,
n^j e_{A}^k \epsilon_{jk \langle i_l}  \widehat n_{ L-1 \rangle}.
\end{equation}

\subsection*{Electric parity tensor harmonics}
The electric parity tensor harmonic is defined by
\be
Y_{AB}^{l m}
:=
\left[
D_A D_B + \frac{1}{2}l(l+1)\,\mathring{q}_{AB}
\right] Y_{l m},
\ee
where $\mathring{q}_{AB}$ is the round metric on the unit two-sphere. The second term ensures
that the tensor is trace-free. We can also write it as 
\be
Y_{AB}^{l m}
= D_{\langle A} D_{B \rangle} Y_{lm},
\ee
The angular brackets on sphere indices denote the symmetric trace-free projection:
\be
T_{\langle AB\rangle}
:=
T_{(AB)} - \frac{1}{2} \mathring{q}_{AB} \,\mathring{q}^{CD} T_{CD}.
\ee
Often we will denote this projection also as $T_{\langle AB\rangle} = \STF_{AB} T_{AB}$. 

To express the electric-parity tensor harmonics in STF form, we begin with
an identity relating $D_B e_A^{\ i}$ to $\mathring{q}_{AB} n^i$.
Since $n^i n_i = 1$, differentiation immediately yields
\begin{equation}
n_i e_A^{i} = 0,
\end{equation}
so that $e_A^{i}$ is tangent to the unit sphere. Differentiating again,
\begin{equation}
D_B(n_i e_A^{i}) = (D_B n_i) e_A^{i} + n_i D_B e_A^{i} = 0.
\end{equation}
This gives,
\begin{equation}
n_i D_B e_A^{i} = -\mathring{q}_{AB}.
\end{equation}
The quantity $D_B e_A^{i}$ is a three-dimensional vector. It can have components in the $n^i$ direction or in the $e^i_C$ directions. Splitting the vector we have,
\be
D_B e_A^{i} = (n_j D_B e_A^{j}) n^i + (e_{j}^C D_B e_A^{j} ) e^i_C
\ee
Expanding the covariant derivative in the second term $e_{j}^C D_B e_A^{j}$, we note that
\bea
e_{j}^C D_B e_A^{j}  &=& e_{j}^C \partial_B e_A^{j} - e_{j}^C \Gamma_{BA}^D e_D^{j} \\
&=& e_{j}^C \partial_B e_A^{j} - \frac{1}{2} e_{j}^C \mathring{q}^{DE} \left(\partial_B \mathring{q}_{AE} + \partial_A \mathring{q}_{BE}  - \partial_E \mathring{q}_{BA}  \right)e_D^{j} = 0,
\eea
where in the last step we use $\mathring{q}_{AB} = e_{Ai} e_{Bi}$ and repeatedly use $\partial_{A} \partial_{B} n^i = \partial_B \partial_A n^i$. This gives us the required identity 
\be \label{eq:De_identity}
D_B e_A^{i} = -\mathring{q}_{AB} n^i.
\ee 
We will make use of this repeatedly.

In the vector harmonics discussion, we already noted that, 
\begin{equation}
D_B \widehat n_{ L}
=
l e_{B \langle i_l} \widehat n_{L-1 \rangle}.
\end{equation}
Applying $D_A$ on this equation, we have,
\begin{align}
D_A D_B \widehat n_{ L}
&=
l \, D_A \bigl( e_{B \langle i_l} \widehat n_{ L-1\rangle} \bigr) \\
&=
l \, D_A e_{ B \langle i_l} \widehat n_{ L-1\rangle}
+ l\,  D_A \widehat n_{\langle L-1} e_{i_l \rangle B} \\
&=
-l \, \mathring{q}_{AB}  n_{\langle i_l} \widehat n_{ L-1\rangle}
+ l(l-1)\,  \STF_L  \left[ e_{A i_{l-1} } \, e_{B i_l}  \, \widehat n_{L-2} \right], \\
&= -l \, \mathring{q}_{AB} \widehat n_{ L} + l(l-1)\,  \STF_L  \left[ e_{A i_{l} } \, e_{B i_{l-1}}  \, \widehat n_{L-2} \right].
\end{align}
Taking the STF part in the $AB$ indices removes the trace term:
\begin{equation}
D_{\langle A} D_{B\rangle} \widehat n_{ L}
=
l(l-1)\,
\STF_{AB} \left[ \STF_L \left[ e_{ A i_l} \, e_{B i_{l-1}}
\, \widehat n_{  L-2} \right] \right].
\end{equation}
We have our final relation, 
\be
Y_{AB}^{lm} =  l(l-1) \mathcal{Y}^{* \langle L \rangle}_{l m} \,
\STF_{AB} \left[ \STF_L \left[ e_{ A i_l} \, e_{B i_{l-1}}
\, \widehat n_{  L-2} \right] \right]. 
\ee

\subsection*{Magnetic parity tensor harmonics}

The magnetic parity tensor harmonic is obtained by applying the Levi--Civita
tensor on the sphere:
\begin{equation}
X_{AB}^{l m}
:=
\epsilon_{(A}{}^{C} D_{B)} D_C Y_{l m} = \STF_{AB} \left[ \epsilon_{A}{}^{C} D_{B} D_C Y_{l m}\right] .
\end{equation}
In terms of the three-dimensional STF Cartesian tensors this becomes
\begin{equation}
X_{AB}^{l m}
=
l(l-1)\,
\mathcal{Y}^{*\langle L \rangle}_{l m}
\STF_{AB} \left[ \STF_{L} \left[ 
\epsilon_{A}{}^{C}   e_{B i_l}
\, e_{C i_{l-1}}
\, \widehat n_{L-2} \right] \right] .
\end{equation}
Equivalently, one may express this in terms of the three-dimensional
Levi--Civita tensor as
\begin{equation}
X_{AB}^{l m}
= l (l-1)
\mathcal{Y}^{*\langle L \rangle}_{l m} \STF_{AB} \left[\STF_{L} \left[ 
\epsilon_{i_l j k}   \, n_j  \,
e_{Ak} \, e_{Bi_{l-1}}
\, \widehat n_{ L-2 }\right]\right].
\end{equation}

\subsection*{Convenient notation}

It is very useful to introduce the following notation 
\bea
Y_A^{l m} &=& \mathcal{Y}^{*\langle L \rangle}_{l m} Y_{A}^{\langle L \rangle} \qquad \mbox{where} \qquad Y_{A}^{\langle L \rangle}  = l \, e_{A \langle i_l} \widehat n_{L-1 \rangle}, \\
X_A^{l m} &=& \mathcal{Y}^{*\langle L \rangle}_{l m} X_{A}^{\langle L \rangle} \qquad \mbox{where} \qquad X_{A}^{\langle L \rangle}  = l \, 
n^j e_{A}^k \epsilon_{jk \langle i_l}  \widehat n_{ L-1 \rangle},
\eea
and
\bea
Y_{AB}^{l m} &=& \mathcal{Y}^{*\langle L \rangle}_{l m} Y^{\langle L \rangle}_{AB}, \\ 
X_{AB}^{l m}
&=& 
\mathcal{Y}^{*\langle L \rangle}_{l m} X^{\langle L \rangle}_{AB}, 
\eea
where
\bea
Y^{\langle L \rangle}_{AB} &:=&  l (l-1)  \STF_{AB} \left[ \STF_L \left[ e_{ A i_l} \, e_{B i_{l-1}}
\, \widehat n_{  L-2} \right] \right], 
\label{electric-L-notation} \\
X^{\langle L \rangle}_{AB} &:=& l (l-1) \STF_{AB} \left[ \STF_{L} \left[
\epsilon_{i_l j k}   \, n_j  \,
e_{Ak} \, e_{Bi_{l-1}}
\, \widehat n_{ L-2 }\right]\right]. \label{magnetic-L-notation}
\eea

\section{Decomposition of \texorpdfstring{$e_A{}_i \widehat{n}_L P_{iL}$}{eA-i nL PiL}}

\label{app:C}

The Damour--Iyer \cite{Damour:1990gj} decomposition of the product of a vector $U_i$ with a symmetric
trace-free (STF) tensor $\widehat{T}_L$ of rank $l$ into irreducible pieces  corresponds to the Clebsch--Gordan decomposition
$D_1 \otimes D_l = D_{l+1} \oplus D_l \oplus D_{l-1}$. The three irreducible
pieces of ranks $l+1$, $l$, and $l-1$ are given explicitly by
\begin{equation}
U_i\,\widehat{T}_L = \widehat{R}^{(+)}_{iL}
+ \frac{l}{l+1}\,\epsilon_{si\langle i_l}\widehat{R}^{(0)}_{L-1\rangle s}
+ \frac{2l-1}{2l+1}\,\delta_{i\langle i_l}\widehat{R}^{(-)}_{L-1\rangle}\,,
\label{vector-decomposition}
\end{equation}
where the three ``brick'' STF tensors are defined as
\begin{align}
& \widehat{R}^{(+)}_{L+1} = U_{\langle i_{l+1}}\widehat{T}_{L\rangle}, \\
& \widehat{R}^{(0)}_{L} = U_a\,\widehat{T}_{b\langle L-1}\, \epsilon_{i_l\rangle ab},  \\
&  \widehat{R}^{(-)}_{L-1} = U_s\,\widehat{T}_{sL-1}\,.
\end{align}
A particularly useful special case of the above decomposition arises when 
\be
U_i = n_i \qquad \mbox{and} \qquad \widehat{T}_L = \widehat{n}_L, 
\ee
for which $\widehat{R}^{(0)}_L = 0$ (since
$\epsilon_{sab}n_a n_b = 0$) and the decomposition reduces to the
\emph{peeling formula}
\begin{equation}
n_i\,\widehat{n}_L = \widehat{n}_{iL}
+ \frac{l}{2l+1}\,\delta_{i\langle i_l}\widehat{n}_{L-1\rangle}. \label{peeling-formula}
\end{equation}
Equation~\eqref{peeling-formula} is used
repeatedly throughout  to ``peel'' $n_i$ out of $\widehat{n}_{iL}$.

Since $D_A(n^i n_i) = 0$, 
it follows that $e_A{}_i = D_A n_i$ (with $A = \theta, \phi$) 
are tangent vectors to the unit sphere\footnote{The index $A$ remains a spectator throughout the following manipulations.}  
$ e_A{}_i\, n^i = 0. $ With
\be
U_i = e_A{}_i \qquad \mbox{and} \qquad  \widehat{T}_L = \widehat{n}_L,
\ee  the 
product $e_A{}_i\widehat{n}_L$ decomposes into three irreducible STF pieces as:
\begin{align} \label{R-plus-final}
& \widehat{R}^{(+)}_{L+1} = e_A{}_{\langle i_{l+1}}\widehat{n}_{L\rangle} = \frac{1}{l+1} Y_A^{\langle L+1 \rangle},  \\
& \widehat{R}^{(0)}_{L} = e_A{}_a\,\widehat{n}_{b\langle L-1}\,\epsilon_{i_l\rangle ab},  \\
&  \widehat{R}^{(-)}_{L-1} = e_A{}_s\,\widehat{n}_{sL-1}\,.
\end{align}
The $\widehat{R}^{(0)}_{L}$ and $\widehat{R}^{(-)}_{L-1}$ pieces are simplified using the peeling formula \eqref{peeling-formula}:
\begin{equation}
\widehat{R}^{(0)}_{L} = e_A{}_a\, n_b \, \widehat{n}_{\langle L-1}\,\epsilon_{i_l\rangle ab} = -\frac{1}{l} X_{A}^{\langle L \rangle}, \label{R-zero-final}
\end{equation} 
\begin{equation}
\widehat{R}^{(-)}_{L-1} = e_{As}\widehat{n}_{sL-1} = -\frac{l-1}{2l-1}\, e_{A\langle i_{l-1}}\widehat{n}_{L-2\rangle} = -\frac{1}{2l-1}\, Y_{A}^{\langle L-1 \rangle} \label{R-minus-final}.
\end{equation}

Let $P_{iL}$ be a general rank-$(l+1)$ tensor. Consider contracting both sides of the decomposition \eqref{vector-decomposition} with $P_{iL}$. At this point, it is useful to recall  that for any two rank-$m$ Cartesian tensors $A_M$ and $B_M$,
\begin{equation}
\widehat{A}_M B_M = A_M \widehat{B}_M = \widehat{A}_M \widehat{B}_M. \label{contraction-lemma}
\end{equation}
The three terms are computed as follows:

\subsubsection*{$\widehat{R}^{(+)}_{ L+1 }$ term}

The first term is simply 
\begin{equation}
\widehat{R}^{(+)}_{iL} \widehat{P}_{iL} = \frac{1}{l+1} Y_{A}^{\langle L+1 \rangle} \widehat P_{\langle L+1\rangle}.
\end{equation} 

\subsubsection*{$\widehat{R}^{(0)}_{ L }$ term}

The second term is:
\bea
\frac{l}{l+1}\,\epsilon_{si\langle i_l}\widehat{R}^{(0)}_{L-1\rangle s} P_{iL} 
&=& - \frac{1}{l+1}\,\epsilon_{si i_l}X_A^{\langle L-1 s\rangle} \STF_L \left[ P_{iL} \right] \\
&=& - \frac{1}{l+1} \,
X_A^{\langle L-1 s\rangle} 
\, \STF_{L-1, s} \left\{ \epsilon_{s i i_l}\left[\STF_L P^{iL}\right] \right\}.
\eea 
The final expression is non-zero for general $P_{iL}$; it vanishes only if $P_{iL}$ is symmetric in all indices.

\subsubsection*{$\widehat{R}^{(-)}_{ L-1}$ term}

For the last term in the decomposition we have
\bea
\frac{2l-1}{2l+1}\,\delta_{i\langle i_l}\widehat{R}^{(-)}_{L-1\rangle} P_{iL} &=& 
- \frac{1}{2l+1}\,\delta_{i i_l}Y_A^{\langle L-1 \rangle } \STF_L P_{iL} \\
&=&- \frac{1}{2l+1} Y_A^{\langle L-1 \rangle }\STF_{L-1}\left[\left[\STF_L P_{iL}\right]^{i_l=i}\right].
\eea

\subsubsection*{Final expression}

Putting everything together, we have our final formula for the decomposition 
\bea
e_A{}_i \widehat{n}_L P_{iL} &=& 
\frac{1}{l+1} Y_{A}^{\langle L+1 \rangle} \widehat P_{\langle L+1\rangle}
- \frac{1}{l+1} \,
X_A^{\langle L-1, s\rangle} 
\, \STF_{L-1, s} \left[ \epsilon_{s i i_l} P^{i \langle L \rangle} \right] \nonumber \\
&& - \frac{1}{2l+1} Y_A^{\langle L-1 \rangle }\STF_{L-1}\left[\delta_{i i_l} P_{i \langle L \rangle}\right]. \label{final_C19}
\eea

\section{Decomposition of \texorpdfstring{$ \STF_{ij} \left[e_{A}{}_i e_{B}{}_j \right]  \widehat{n}_L S_{ijL}  $}{STF(eAeB nL SijL)}}

\label{app:D}

The tensor product of a
symmetric rank-2 tensor $S_{ij}$ with an STF tensor $\widehat{T}_L$ of rank $l$ decomposes
into six irreducible pieces~\cite{Damour:1990gj}:
\begin{align}
S_{ij}\widehat{T}_L 
&= H_{ijL}^{(+2)}  \nonumber \\ & 
+ \mathrm{STF}_{ij}\!\Bigl[
\epsilon_{ai\langle i_l}\,H^{(+1)}_{L-1\rangle j a}
+ \delta_{i\langle i_l}\,H^{(0)}_{L-1\rangle j }
+ \epsilon_{a i\langle i_l}\,H^{(-1)}_{L-2 |a|} \delta_{i_{l-1} \rangle j}
+ \delta_{i\langle i_l}\,H^{(-2)}_{L-2}\,\delta_{i_{l-1}\rangle j}
\Bigr] \nonumber \\ &  + \delta_{ij}K_L \,, \label{eq:DI_2.11}
\end{align}
where the six STF building blocks are defined in terms of $S_{ij}$ and $\widehat{T}_L$ as
\begin{align}
&  H^{(+2)}_{L+2} = \widehat{S}_{\langle i_{l+2} i_{l+1}}\widehat{T}_{L\rangle} \,,
\label{eq:DI_2.12a}\\[3pt]
& H^{(+1)}_{L+1} = \frac{2l}{l+2}\, \widehat{S}_{c\langle i_l}\,\widehat{T}_{|d|L-1}\,\epsilon_{i_{l+1} \rangle c d} \,,
\label{eq:DI_2.12b}\\[3pt]
&  H^{(0)}_{L}     = \frac{6l(2l-1)}{(l+1)(2l+3)}\,
\widehat{S}_{a\langle i_l}\widehat{T}_{L-1\rangle a} \,,
\label{eq:DI_2.12c}\\[3pt]
& H^{(-1)}_{L-1}  = \frac{2(l-1)(2l-1)}{(l+1)(2l+1)}\,
\widehat{S}_{ca}\,\widehat{T}_{bc \langle L-2}\,\epsilon_{i_{l-1}\rangle ab} \,,
\label{eq:DI_2.12d}\\[3pt]
&  H^{(-2)}_{L-2}  = \frac{2l-3}{2l+1}\,\widehat{S}_{ab}\widehat{T}_{abL-2} \,,
\label{eq:DI_2.12e}\\[3pt]
&  K_{L}           = \tfrac{1}{3}\,S_{aa}\widehat{T}_{L} \,.
\label{eq:DI_2.12f}
\end{align}
For $S_{ij} = n_i n_j$ and $\widehat{T}_L = \widehat{n}_L$, the above decomposition yields
\bea
\label{eq:DI_2.13}
n_a n_b\,\widehat{n}_L 
& =& \widehat{n}_{abL}
+ \frac{l}{2l+3}\!\left(
\widehat{n}_{a\langle L-1}\delta_{i_l\rangle b}
+ \widehat{n}_{b\langle L-1}\delta_{i_l\rangle a}
\right)
+ \frac{\delta_{ab}}{2l+3}\,\widehat{n}_{L} \nonumber \\ &&
+ \frac{l(l-1)}{(2l+1)(2l-1)}\,
\widehat{n}_{\langle L-2}\,\delta_{i_{l-1}|a|}\,\delta_{i_l\rangle b} \,.
\eea

We want to take $S_{ij} =  \STF_{ij} e_{Ai} e_{Bj}$ and $\widehat{T}_L = \widehat{n}_L$, where $A$ and $B$ are spectator indices. Since the $S_{ij}$ tensor is trace-free the $K_L$ term is zero. There are five terms in such a decomposition. These are:
\subsubsection*{$H^{(+2)}_{L+2}$ term}  
$H^{(+2)}_{L+2}$ term is 
\be
H^{(+2)}_{L+2} =  \STF_{L+2}  \left[ e_{A i_{l+2}} e_{B i_{l+1}} \widehat n_{L}\right],
\ee
which can be written as
\bea
H^{(+2)}_{L+2} & = &  \STF_{AB}  \STF_{L+2}  \left[ e_{A i_{l+2}} e_{B i_{l+1}} \widehat n_{L}\right] - \frac{1}{2} \mathring{q}_{AB} \widehat{n}_{L+2} \\
&=&
\frac{1}{(l+2)(l+1)} Y_{AB}^{\langle L+2 \rangle} - \frac{1}{2} \mathring{q}_{AB} \widehat{n}_{L+2}.
\eea
\subsubsection*{$H^{(+1)}_{L+1}$ term}  
To work out the $H^{(+1)}_{L+1}$ term, let us recall the magnetic tensor harmonic notation introduced in Eq.~\eqref{magnetic-L-notation},
\be
X^{\langle L+1 \rangle}_{AB} = l (l+ 1) \STF_{AB} \left[ \STF_{L+1} \left[
\epsilon_{i_{l+1} j k}   \, n_j  \,
e_{Ak} \, e_{Bi_{l}}
\, \widehat n_{ L-1 }\right]\right].
\ee
We also need the peeling formula \eqref{peeling-formula} with $L \to L-1$:
\begin{equation} 
\widehat{n}_{L-1}^d  = n^d\,\widehat{n}_{L-1}
- \frac{l-1}{2l-1}\,\delta^d_{\langle i_{l-1}}\widehat{n}_{L-2\rangle}. \end{equation}
We have, 
\bea
H^{(+1)}_{L+1} &=& \frac{2l}{l+2}\, \STF_{L+1} \left[ \STF_{c i_l} \left[ e_{cA} e_{i_l B} \right]\,\widehat{n}^d_{L-1}\,\epsilon_{i_{l+1}  c d} \right] \\
&=& 	- \frac{2}{(l+1)(l+2)} X^{\langle L+1 \rangle}_{AB} .
\eea

\subsubsection*{$H^{(0)}_{L}$ term}  
$H^{(0)}_{L}$ term is 
\be
H^{(0)}_{L} = \frac{6l(2l-1)}{(l+1)(2l+3)}\,
\STF_{L} \left[ \STF_{a i_l} \left[ e_{Aa} e_{B i_l} \right]  \widehat{n}_{L-1 a} \right]. 
\ee
To simplify this we need to use the peeling formula \eqref{peeling-formula}. We get, 
\bea
H^{(0)}_{L} &=& \frac{6l(2l-1)}{(l+1)(2l+3)}\,
\STF_{L} \left[ \STF_{a i_l} \left[ e_{Aa} e_{B i_l} \right] \left( \widehat{n}_{L-1} n_a - \frac{l-1}{2l-1} \delta_{a i_{l-1}} \widehat{n}_{L-2}\right) \right] \nonumber \\
&=& \frac{6l(2l-1)}{(l+1)(2l+3)} \left\{- \frac{l-1}{2l-1} \STF_{L}  \left[ e_{A i_{l}} e_{B i_{l-1}} \widehat n_{L-2}\right] - \frac{1}{3} \mathring{q}_{AB} \widehat{n}_{L} \right\} \\
&=& -\frac{6}{(l+1)(2l+3)} 
Y_{AB}^{\langle L \rangle} - \frac{l}{2l+3} \mathring{q}_{AB} \widehat{n}_{L}.
\eea
\subsubsection*{$H^{(-1)}_{L-1}$ term}  
The $H^{(-1)}_{L-1}$ term is
\be
H^{(-1)}_{L-1}  = \frac{2(l-1)(2l-1)}{(l+1)(2l+1)}\,
\STF_{ca} \left[ e_{cA} e_{aB} \right] \,\widehat{n}_{bc \langle L-2}\,\epsilon_{i_{l-1}\rangle ab}.
\ee
Using \eqref{eq:DI_2.13} to expand out $\widehat{n}_{bcL-2}$, we find, 
\begin{align}
\widehat{n}_{bcL-2} &= n_b n_c \widehat{n}_{L-2}
- \frac{l-2}{2l-1}\!\left(\widehat{n}_{b\langle L-3}\,\delta_{i_{l-2}\rangle c}
+ \widehat{n}_{c\langle L-3}\,\delta_{i_{l-2}\rangle b}\right) \notag\\
&\quad - \frac{\delta_{bc}}{2l-1}\,\widehat{n}_{L-2}
- \frac{(l-2)(l-3)}{(2l-3)(2l-5)}\,\widehat{n}_{\langle L-4}\,
\delta_{ i_{l-3}}^{b}\,\delta_{i_{l-2}\rangle}^{c} \label{eq:inversion-another}.
\end{align}
To simplify it for our use we also need the \eqref{peeling-formula} with $L\to L-3$:
\begin{equation} 
\widehat{n}_{bL-3}  = n_b\,\widehat{n}_{L-3}
- \frac{l-3}{2l-5}\,\delta_{b \langle i_{l-3}}\widehat{n}_{L-4\rangle}. \end{equation}
We have the definition 
\be
X^{\langle L-1 \rangle}_{AB} = (l- 1) (l- 2) \STF_{AB} \left[ \STF_{L-1} \left[
\epsilon_{i_{l-1} j k}   \, n_j  \,
e_{Ak} \, e_{Bi_{l-2}}
\, \widehat n_{ L-1 }\right]\right],
\ee
using which we get
\be
H^{(-1)}_{L-1}  = \frac{2}{(l+1)(2l+1)}X^{\langle L-1 \rangle}_{AB}.
\ee

\subsubsection*{$H^{(-2)}_{L-2}$ term}  
$H^{(-2)}_{L-2}$ term is 
\be
H^{(-2)}_{L-2}  = \frac{2l-3}{2l+1}\,\STF_{ab} \left[e_{Aa} e_{Bb} \right] \widehat{n}_{abL-2}. \label{H-2-L-2}
\ee
Using \eqref{eq:DI_2.13} to expand out $\widehat{n}_{abL-2}$, we find, 
\begin{align}
\widehat{n}_{abL-2} &= n_a n_b \widehat{n}_{L-2}
- \frac{l-2}{2l-1}\!\left(\widehat{n}_{a\langle L-3}\,\delta_{i_{l-2}\rangle b}
+ \widehat{n}_{b\langle L-3}\,\delta_{i_{l-2}\rangle a}\right) \notag\\
&\quad - \frac{\delta_{ab}}{2l-1}\,\widehat{n}_{L-2}
- \frac{(l-2)(l-3)}{(2l-3)(2l-5)}\,\widehat{n}_{\langle L-4}\,
\delta_{ i_{l-3}}^{a}\,\delta_{i_{l-2}\rangle}^{b} \label{eq:inversion}.
\end{align}
To simplify it for our use we also need the \eqref{peeling-formula} with $L\to L-3$:
\begin{equation} \label{simple-peeling}
\widehat{n}_{aL-3}  = n_a\,\widehat{n}_{L-3}
- \frac{l-3}{2l-5}\,\delta_{a \langle i_{l-3}}\widehat{n}_{L-4\rangle}. \end{equation}
Inserting \eqref{eq:inversion} and \eqref{simple-peeling} in \eqref{H-2-L-2}  gives us 
\bea H^{(-2)}_{L-2}  &=& - \frac{(2l-3)}{(2l+1)(2l-1)} \mathring{q}_{AB} \widehat{n}_{L-2} +  \frac{(l-2)(l-3)}{(2l+1)(2l-1)} \STF_{L-2} \left[ e_{A i_{l-2}} e_{B i_{l-3}} \widehat{n}_{L-4} \right] \\
&=& - \frac{l(l-1)}{2(2l+1)(2l-1)} \mathring{q}_{AB} \widehat{n}_{L-2} +  \frac{(l-2)(l-3)}{(2l+1)(2l-1)} \STF_{AB} \STF_{L-2} \left[ e_{A i_{l-2}} e_{B i_{l-3}} \widehat{n}_{L-4} \right] \nonumber \\
&=& - \frac{l(l-1)}{2(2l+1)(2l-1)} \mathring{q}_{AB} \widehat{n}_{L-2} +  \frac{1}{(2l+1)(2l-1)} Y_{AB}^{ \langle L-2 \rangle}.
\eea

\subsubsection*{Contraction with $S_{ijL}$}
Now we want to contract these terms with $S_{ijL}$. The first term is 
\be
\left\{ \frac{1}{(l+2)(l+1)} Y_{AB}^{\langle L+2 \rangle} - \frac{1}{2} \mathring{q}_{AB} \widehat{n}_{L+2} \right\} S_{\langle L+2 \rangle}. 
\ee
The second term  is
\be
- \frac{2}{(l+1)(l+2)} X_{AB}^{\langle L-1, j, a \rangle } \,  \STF_{L-1, j, a } \left[  \epsilon_{ai i_l} S_{\langle ij \rangle \langle L \rangle} \right]. \label{final_D30}  
\ee
The third term  is
\be
\left\{ - \frac{6}{(l+1)(2l+3)}Y_{AB}^{\langle L-1, j \rangle} - \frac{l}{2l+ 3} \mathring q_{AB} \widehat{n}^{L-1,j} \right\}  \,  \STF_{L-1,j} \left[ \delta_{i i_l} S_{\langle ij \rangle \langle L \rangle} \right].
\label{final_D31}
\ee
The fourth term is
\be
\frac{2}{(l+1)(2l+1)}X_{AB}^{\langle L-2, a \rangle} \, \STF_{L-2,a} \left[ \epsilon_{a i i_l}  \delta_{i_{l-1}  j} S_{\langle ij \rangle \langle L \rangle} \right],
\ee
and the last term is 
\be
\left\{ - \frac{l(l-1)}{2(2l+1)(2l-1)} \mathring{q}_{AB} \widehat{n}_{L-2} +  \frac{1}{(2l+1)(2l-1)} Y_{AB}^{ \langle L-2 \rangle} \right\} \STF_{L-2} \left[ \delta_{i i_l} \, \delta_{i_{l-1} j} S_{\langle ij \rangle \langle L \rangle} \right].
\ee
Our final formula for the decomposition of $ \STF_{ij} \left[e_{A}{}_i e_{B}{}_j \right]  \widehat{n}_L S_{ijL}  $ is the sum of the above five terms. 

\section{STF tensors in \texttt{Mathematica}} 
\label{STF-Mathematica}

Readers not interested in the implementation of the above calculations in 
\texttt{Mathematica} can ignore this appendix. This appendix documents the core \texttt{Mathematica} functions underlying the STF tensor manipulations presented in the main text. We follow the conventions of Blanchet and Damour~\cite{Blanchet:1985sp} and Damour and Iyer~\cite{Damour:1990gj} throughout. The selection below is not exhaustive but captures the essential structure of the implementation; readers familiar with STF methods will recognise that related functionality exists in other \texttt{Mathematica} packages as well in \texttt{Mathematica}.  We found it useful to write our own functions.

\subsection*{\texttt{tensorProd}}
\hrule

\medskip

\texttt{tensorProd[A, B]} forms the tensor outer product of \texttt{A}
and \texttt{B}, falling back to ordinary scalar multiplication if either
argument is not an array. The variation \texttt{tensorProd[A, B, rest\_\_]} chains multiple products via
\texttt{Fold}.

\medskip
\noindent
\texttt{tensorProd[A\_, B\_] :=}\\
\texttt{\quad Which[ArrayQ[A] \&\& ArrayQ[B],  TensorProduct[A, B],}\\
\texttt{\quad\quad\quad ArrayQ[A] \&\& !ArrayQ[B], B A,}\\
\texttt{\quad\quad\quad !ArrayQ[A] \&\& ArrayQ[B], A B, True, A B]}

\medskip
\noindent
\texttt{tensorProd[A\_, B\_, rest\_\_] :=}\\
\texttt{\quad Fold[tensorProd, tensorProd[A, B], \{rest\}]}

\medskip
\hrule
\bigskip

\subsection*{\texttt{symmetrizeTensor}}
\hrule
\medskip
\texttt{symmetrizeTensor[T]} symmetrizes a rank-$r$ tensor by averaging
over all $r!$ permutations of its indices,
\be
T_{(i_1\cdots i_r)} = \frac{1}{r!}\sum_{\sigma\in S_r}
T_{i_{\sigma(1)}\cdots i_{\sigma(r)}},
\ee
implemented explicitly via \texttt{Transpose}.

\medskip

\medskip
\noindent \texttt{symmetrizeTensor[T\_] :=}\\
\texttt{\quad Module[\{r = ArrayDepth[T], perms\},}\\
\texttt{\quad\quad perms = Permutations[Range[r]];} \\
\texttt{\quad\quad Total[Transpose[T, \#] \& /@ perms] / Length[perms]]}

\medskip
\noindent Although equivalent to Mathematica's built-in \texttt{Symmetrize[T]}, this explicit implementation was retained as an independent cross-check.

\medskip
\hrule
\bigskip

\subsection*{\texttt{tensor}}
\hrule
\medskip

\texttt{tensor[sym, n]} allocates a symbolic rank-$n$ tensor in three
dimensions, returning a $3^n$ array whose $(i_1,\ldots,i_n)$ entry is
\texttt{sym[}$i_1$\texttt{,}\ldots\texttt{,}$i_n$\texttt{]}.

\medskip
\noindent \texttt{tensor[sym\_, n\_Integer?Positive] :=}
\texttt{\quad Array[sym, ConstantArray[3, n]]}

\medskip
\hrule
\bigskip

\subsection*{\texttt{deltaOuter}}
\hrule
\medskip

\texttt{deltaOuter[n]} constructs the outer product of $n$ copies of
the $3\times 3$ identity matrix, yielding the rank-$2n$ tensor
$\delta_{i_1 j_1}\otimes\cdots\otimes\delta_{i_n j_n}$.

\medskip
\noindent \texttt{deltaOuter[n\_Integer?Positive] :=}\\
\texttt{\quad TensorProduct @@ ConstantArray[IdentityMatrix[3], n]}

\medskip
\hrule
\bigskip

\subsection*{\texttt{traceLast2k}}
\hrule
\medskip

\texttt{traceLast2k[T, k]} contracts the last $2k$ indices of \texttt{T}
in consecutive pairs, implementing $k$ successive traces
\be
T_{ i_{2k+1}\cdots i_r m_1 m_1 m_2 m_2 \cdots m_k m_k}.
\ee
The function returns \texttt{\$Failed} if $2k > r$.

\medskip
\noindent \texttt{traceLast2k[T\_, k\_Integer?NonNegative] :=}\\
\texttt{\quad Module[\{r = ArrayDepth[T], pairs\},}\\
\texttt{\quad\quad If[2k > r, Return[\$Failed]];} \\
\texttt{\quad\quad pairs = Table[\{r-2k+2j-1, r-2k+2j\}, \{j, 1, k\}];} \\
\texttt{\quad\quad TensorContract[T, pairs]]}
\medskip

\hrule
\bigskip

\subsection*{\texttt{acoeff}}
\hrule
\medskip

\texttt{acoeff[p, k]} returns the rational coefficient
\[
a(p,k) = \frac{(-1)^k\, p!}{(2p-1)!!}
\frac{(2p-2k-1)!!}{(p-2k)! \,(2k)!!}
\]
that appears in the STF projection, ensuring exact trace cancellation
at each order $k$.

\medskip
\noindent \texttt{acoeff[p\_, k\_] :=}\\
\texttt{\quad ((-1)\^{}k p!) / (p-2k)! * (2p-2k-1)!! / ((2p-1)!! (2k)!!)}

\medskip
\hrule
\bigskip

\subsection*{\texttt{partialTrace}}
\hrule
\medskip

\texttt{partialTrace[T, inds]} contracts \texttt{T} over the index
positions listed in \texttt{inds}, grouping them into consecutive pairs.
Validates that \texttt{inds} has even length, contains integers with no
duplicates, and that all positions lie within the rank of \texttt{T}.

\medskip
\noindent \texttt{partialTrace[T\_, inds\_List] :=}\\
\texttt{\quad Module[\{r = ArrayDepth[T], pairs\},}\\
\texttt{\quad\quad If[OddQ[Length[inds]], Return[\$Failed]];} \\
\texttt{\quad\quad If[!VectorQ[inds, IntegerQ], Return[\$Failed]];} \\
\texttt{\quad\quad If[Length[DeleteDuplicates[inds]] < Length[inds], Return[\$Failed]];} \\
\texttt{\quad\quad If[Min[inds] < 1 || Max[inds] > r, Return[\$Failed]];} \\
\texttt{\quad\quad pairs = Partition[inds, 2];} \\
\texttt{\quad\quad TensorContract[T, pairs]]}

\medskip
\noindent For example, on a rank-8 tensor $T$ \texttt{partialTrace[T, \{2,3,5,6\}]} would give a rank-4 tensor
\be
T_{i_1 i_2 i_2 i_4 i_5 i_5 i_7 i_8}. 
\ee

\medskip
\hrule
\bigskip

\subsection*{\texttt{tensorDotIndices}}
\hrule
\medskip

\texttt{tensorDotIndices[A, B, setA, setB]} contracts tensor \texttt{A}
against tensor \texttt{B} by pairing the indices in \texttt{setA} with
those in \texttt{setB},
\be
\sum_{m_1,\ldots,m_k}
A_{\cdots m_1\cdots m_k\cdots}\,B_{\cdots m_1\cdots m_k\cdots}.
\ee
Since index $j$ of \texttt{B} sits at position $r_A+j$ in
\texttt{TensorProduct[A,B]}, the contraction pairs are built as
\texttt{Transpose[\{setA, rA + setB\}]}.

\medskip
\noindent \texttt{tensorDotIndices[A\_, B\_, setA\_List, setB\_List] :=}\\
\texttt{\quad Module[\{rA, rB, pairs\}, rA = ArrayDepth[A]; rB = ArrayDepth[B];}\\
\texttt{\quad\quad If[Length[setA] =!= Length[setB], Return[\$Failed]];} \\
\texttt{\quad\quad If[!VectorQ[setA,IntegerQ]||!VectorQ[setB,IntegerQ], Return[\$Failed]];} \\
\texttt{\quad\quad If[Length[DeleteDuplicates[setA]]<Length[setA], Return[\$Failed]];} \\
\texttt{\quad\quad If[Length[DeleteDuplicates[setB]]<Length[setB], Return[\$Failed]];} \\
\texttt{\quad\quad If[Min[setA]<1||Max[setA]>rA, Return[\$Failed]];} \\
\texttt{\quad\quad If[Min[setB]<1||Max[setB]>rB, Return[\$Failed]];} \\
\texttt{\quad\quad pairs = Transpose[\{setA, rA + setB\}];} \\
\texttt{\quad\quad TensorContract[TensorProduct[A, B], pairs]]}

\medskip
\hrule
\bigskip

\subsection*{\texttt{STF}}
\hrule
\medskip

\texttt{STF[Tensor]} computes the Symmetric Trace-Free projection over
all indices via the explicit formula
\[
\widehat{T}_{i_1\cdots i_l} =
\sum_{k=0}^{\lfloor l/2\rfloor} a(l,k)\,
\delta_{(i_1 i_2}\cdots\delta_{i_{2k-1}i_{2k}}\,
T_{i_{2k+1}\cdots i_l) i_1 i_1 i_2 i_2 \cdots i_k i_k},
\]
where the $k=0$ term is \texttt{Symmetrize[Tensor]} and each $k\geq 1$
term subtracts the $k$-th trace of the symmetrized tensor, weighted by
$k$ Kronecker deltas and the coefficient $a(l,k)$.

\medskip
\noindent \texttt{STF[Tensor\_] :=}\\
\texttt{\quad Symmetrize[Tensor] +}\\
\texttt{\quad Sum[acoeff[ArrayDepth[Tensor], k]}\\
\texttt{\quad\quad Symmetrize[tensorProd[deltaOuter[k],}\\
\texttt{\quad\quad\quad traceLast2k[Symmetrize[Tensor], k]]],}\\
\texttt{\quad\quad \{k, 1, Floor[ArrayDepth[Tensor]/2]\}] // Normal}

\medskip
\hrule

\bibliography{dSgrav}
\bibliographystyle{JHEP}

\end{document}